\documentclass[twocolumn,apj,numberedappendix,appendixfloats,twocolappendix]{openjournal}
\usepackage{natbib}

\usepackage{graphicx,amsmath,amssymb,amstext}
\usepackage{amsbsy,amsfonts,amsthm,color}
\usepackage[colorlinks,linkcolor=blue,citecolor=blue,urlcolor=blue ]{hyperref}
\usepackage[utf8]{inputenc}
\usepackage{float}
\usepackage{lipsum}
\usepackage{verbatim}
\usepackage[normalem]{ulem}
\usepackage{orcidlink}
\usepackage{soul}
\usepackage{subcaption}

\begin{document}

\title{SN 2025adpq: A Type Ia supernova in a collisional ring formed during a major galaxy merger}

\author{Brendan O'Connor$^{1,*,\dagger}$\orcidlink{0000-0002-9700-0036}}
\author{Xander J. Hall$^{1}$\orcidlink{0000-0002-9364-5419}}
\author{Tom\'as Cabrera$^{1}$\orcidlink{0000-0002-1270-7666}}
\author{Lei Hu$^{2}$\orcidlink{0000-0001-7201-1938}}
\author{Antonella Palmese$^{1}$\orcidlink{0000-0002-6011-0530}}
\author{Louis-Gregory Strolger$^{3,4}$\orcidlink{0000-0002-7756-4440}}

%Alphabetical
\author{Ariel J. Amsellem$^{1}$\orcidlink{0000-0003-3433-2698}}
\author{Akash Anumarlapudi$^{5}$\orcidlink{0000-0002-8935-9882}}
\author{Igor Andreoni$^{5}$\orcidlink{0000-0002-8977-1498}}
\author{Saul Baltasar$^{6,7}$\orcidlink{0009-0003-4697-7079}}
\author{Jonathan Carney$^{5}$\orcidlink{0000-0001-8544-584X}}
\author{David A. Coulter$^{4,3}$\orcidlink{0000-0003-4263-2228}}
\author{James Freeburn$^{8,9}$\orcidlink{0009-0006-7990-0547}}
\author{Julius Gassert$^{10,1}$\orcidlink{0009-0008-2754-1946}}
\author{Xiaosheng Huang$^{11,6}$\orcidlink{0000-0001-8156-0330}}
\author{Keerthi Kunnumkai$^{1}$\orcidlink{0009-0000-4830-1484}}
\author{Justin D. R. Pierel$^{3}$\orcidlink{0000-0002-2361-7201}}
\author{Mathew R. Siebert$^{3}$\orcidlink{0000-0003-2445-3891}}
\author{Christopher J. Storfer$^{12,6}$\orcidlink{0000-0002-0385-0014}}

\affiliation{$^{1}$McWilliams Center for Cosmology and Astrophysics, Department of Physics, Carnegie Mellon University, Pittsburgh, PA 15213, USA}
\affiliation{$^{\dagger}$McWilliams Fellow}
\affiliation{$^{2}$Department of Physics and Astronomy, University
of Pennsylvania 209 South 33rd Street, Philadelphia, Pennsylvania 19104, USA}
\affiliation{$^{3}$Space Telescope Science Institute, 3700 San Martin Drive, Baltimore, MD 21218, USA}
\affiliation{$^{4}$William H. Miller III Department of Physics and Astronomy, Johns Hopkins University, 3400 North Charles Street, Baltimore, MD 21218, USA}
\affiliation{$^{5}$University of North Carolina at Chapel Hill, 120 E. Cameron Ave., Chapel Hill, NC 27514, USA}
\affiliation{$^{6}$Physics Division, Lawrence Berkeley National Laboratory, 1 Cyclotron Road, Berkeley, CA 94720, USA}
\affiliation{$^{7}$Department of Physics, University of Illinois at Urbana-Champaign, Urbana, IL 61801, USA}
\affiliation{$^{8}$Centre for Astrophysics and Supercomputing, Swinburne University of Technology, Hawthorn, 3122, VIC, Australia}
\affiliation{$^{9}$ARC Centre of Excellence for Gravitational Wave Discovery}
\affiliation{$^{10}$University Observatory, Faculty of Physics, Ludwig-Maximilians-Universität München, Scheinerstr. 1, 81679 Munich, Germany}
\affiliation{$^{11}$Department of Physics \& Astronomy, University of San Francisco, San Francisco, CA 94117, USA}
\affiliation{$^{12}$Institute for Astronomy, University of Hawaii, Honolulu, HI 96822-1897, USA}

\email{$^*$email: boconno2@andrew.cmu.edu}

\begin{abstract}

Galaxy mergers can both trigger star formation and rearrange where stars live, producing long-lived tidal structures and collisionally driven density waves (known as collisional rings) that can extend for tens of kpc from their host galaxy centers. Here we report the discovery of SN 2025adpq, a Type Ia supernova at $z=0.1540$, found within a collisional ring, which we call Pika's Halo, with circumference $\sim$\,70 kpc that was produced by a major merger between two comparable mass galaxies ($\log(M_*/M_\odot)\approx10.5)$. The supernova lies along the ring at a projected offset of $\sim$11.4 kpc from the nucleus of the primary galaxy (hereafter G1). Optical spectroscopy obtained with the Southern African Large Telescope (SALT) and Gemini South reveal signatures consistent with merger induced ongoing star formation, while prominent Calcium H and K absorption indicates a substantial old stellar population within the ring. Therefore, we propose that SN 2025adpq may have been produced by an old progenitor system that was displaced from G1 during the head-on encounter. In this scenario, the progenitor was stripped from its parent galaxy by the collisionally induced pressure wave and exploded far from its birthplace. However, given the broad diversity in SN Ia delay times, we cannot conclusively demonstrate that the progenitor was not formed in a more recent burst of star formation  triggered by the expanding pressure wave. Regardless of the formation pathway, SN 2025adpq highlights collisional rings as a path to producing large offset SNe Ia, and it motivates targeted searches for faint, dynamically displaced old populations in seemingly hostless SN Ia environments. We additionally identify other supernovae, including supernova siblings, in the low redshift sample of collisional ring galaxies, and find that SN 2025adpq is one of only a handful of classified supernova identified in the expanding ring of a collisional ring complex.

\end{abstract}


\section{Introduction}

Interactions and mergers play a fundamental role in galaxy evolution, simultaneously triggering star formation, redistributing stellar populations, and generating tidal structures that influence the galaxy's morphology. One particularly striking outcome of near head-on encounters is the formation of collisional ring galaxies (CRGs), in which a collisionally driven density wave propagates through a disk galaxy \citep{LyndsToomre1976,StruckMarcellLotan1990}, producing expanding rings that can extend for tens of kiloparsecs from the host nucleus \citep[see the reviews by][]{Athanassoula1985,AppletonStruckMarcell1996}. Such systems provide a unique laboratory for studying the interplay between impulsive gravitational perturbations, star formation, and the dynamical displacement of pre-existing stellar populations \citep[see, e.g.,][]{LyndsToomre1976,AppletonStruckMarcell1987,AppletonStruckMarcell1996,HernquistWeil1993}. Historically, these systems were mainly identified in the local Universe at low redshifts \citep{Arp1966,ArpMadore1987,TheysSpiegel1976,Madore2009,Buta2017,Dalcanton2025}. However, with modern deep imaging and spectroscopy, ring-like morphologies are now identified across a wide redshift range, including at high redshift ($z$\,$>$\,$1$\,$-$\,$4$), underscoring their importance as probes of galaxy assembly under extreme dynamical conditions \citep[e.g.,][]{Yuan2020ring,vanDokkum2025ring,Li2025doublering-owl,Khoram2025ring,NestorShachar2025ring,Perna2025ring,Vizgan2026ring}. 

We note that collisional ring galaxies differ from other ring structures such as polar ring galaxies \citep[e.g.,][]{Schweizer1983,Whitmore1990} or resonance rings. CRGs are noticeably more rare (roughly 1\% of all rings; see, e.g., \citealt{Madore2009}) in larger non-uniform samples of ring galaxies \citep[e.g.,][]{Buta2017}, likely due to the requirement of a cataclysmic, bullseye collision. Methods for determining CRGs, and ring galaxies in general, have ranged from visual inspection (the most common method; e.g., \citealt{ArpMadore1987,Madore2009}), citizen science projects \citep[e.g.,][]{Moiseev2011,Buta2017}, algorithmic image analysis \citep[e.g.][]{Timmis2017,Shamir2020}, and machine learning \citep[e.g.,][]{Krishnakumar2024,Zhang2025ringsLS,Dobrycheva2025,Abraham2025}. The latter methods are still not as effective (or pure) as visual inspection, so these works are unlikely to include a complete accounting of all collisional ring galaxies and can be prone to misidentification.

The physical impact of an interaction depends sensitively on encounter geometry (e.g., inclination), mass ratio, and interaction stage \citep[e.g.,][]{LyndsToomre1976,Toomre1978,StruckMarcellLotan1990,Struck2010}. Observationally, interacting systems are commonly classified using projected separation, relative velocity, and visual signs of morphological disturbance (e.g., rings or tails). They are also commonly separated into either major or minor mergers based on their luminosity or stellar mass ratios \citep{Woods2006,Lambas2012,Guzman-Ortega2023}. Numerical simulations show that the strongest features, including tidal tails and collisional rings, arise in major or near equal mass encounters, particularly during the brief merging phase \citep{Madore2009,Wu2012ring,Fiacconi2012,Elagali2018eagle}. This stage is expected to last for only $\lesssim$0.5 Gyr \citep{Wong2006,Lotz2008,Conn2011,Smith2012,Renaud2018,Martinez-Delgado2023}, during which the propagating density wave, and sometimes the host galaxy \citep[e.g.,][]{Renaud2018,Khoram2025ring}, experiences enhanced star formation \citep[e.g.,][]{AppletonStruckMarcell1987,AppletonStruckMarcell1996,DiMatteo2007,DiMatteo2008,Elagali2018sfr}. % but limited production of long delay transient populations.

Supernovae offer an independent diagnostic of these environments, as their rates and spatial distributions reflect the ages of their progenitor systems. During the relatively short merging phase ($\lesssim$0.5 Gyr), galaxies do not have sufficient time to produce large numbers of Type Ia supernovae, whose progenitors have a broad range of delay times \citep{Totani2008,Maoz2012,Maoz2012rev}, while core-collapse supernovae (CC SNe), with progenitor lifetimes $\lesssim$0.1 Gyr, can be produced more efficiently. Observationally, this picture is supported by the enhanced CC SN rates and central excesses observed in strongly disturbed and merging galaxies \citep{Habergham2010,Habergham2012}. %In contrast, Type Ia supernovae preferentially trace older stellar populations \citep{vandenBergh1991,vandenBergh2005,Anderson2015}, though faster formation channels do exist \citep{Childress2014,Wiseman2022}.

Large statistical studies of supernova host galaxies reinforce this interpretation, demonstrating clear differences in the relative frequencies of CC and Type Ia supernovae across host morphologies, star formation properties, and interaction stages \citep[e.g.,][]{Anderson2012,Hakobyan2012,Hakobyan2014}. While interacting and peculiar galaxies show elevated CC SN rates \citep[e.g.,][]{Neff2004,Habergham2010,Habergham2012,Michalowski2020}, Type Ia supernovae remain closely linked to the underlying stellar mass \citep{vandenBergh1991,vandenBergh2005,Anderson2015}. Consequently, Type Ia supernovae discovered at large galactocentric offsets or within tidal and collisional structures \citep[e.g.,][]{SN2007sr,iPTF16abc} provide compelling evidence for the redistribution of old progenitor systems by galaxy interactions \citep[see, e.g.,][]{Ditrani2024}. In this context, supernovae occurring in collisional rings offer a powerful probe of how violent encounters and mergers can displace stellar populations far from their original birth sites.

In this paper we report the discovery of SN 2025adpq, a Type Ia supernova at $z=0.1540$ discovered in a collisional ring, which we refer to as \textit{Pika's Halo}, produced by a major merger between two nearly equal mass galaxies ($\log(M_*/M_\odot)\approx10.5$). The supernova occurred in the ring of radius 10.5 kpc at a projected offset of $\sim$11.4 kpc from the primary galaxy, and spectroscopy indicates both merger driven ongoing star formation and a substantial old stellar population within the ring. We posit that SN 2025adpq originated from an old progenitor system that was dynamically displaced during the head-on encounter, motivating targeted searches for faint, redistributed stellar populations in seemingly ``hostless'' SN Ia environments \citep[e.g.,][]{Strolger2025}.

\section{Observations}

\subsection{Discovery}

We performed follow-up of the gravitational wave event S251112cm with the Dark Energy Camera (DECam) starting on 2025-11-14 at 00:03 UT \citep{2025GCN.42691....1H} as part of the Gravitational Wave MultiMessenger DECam Survey (GW-MMADS; Proposal ID 2023B-851374, PI: Andreoni \& Palmese). The transient SN 2025adpq\footnote{\url{https://www.wis-tns.org/object/2025adpq}} was discovered on 2025-11-14 at 01:47 UT at an apparent $g$-band magnitude of 20.5 AB mag. It was found 250\arcsec\, (66 kpc) from the low redshift HyperLEDA galaxy 2dFGRS TGS493Z081 at $z=0.0125$ (56 Mpc). However, the source lay on top of a ring-like structure near two point-like sources (Figure \ref{fig:Discovery}), and its origin was not clear. Initially it was suspected that the ring-like structure (hereafter referred to as an arc) was a lensed background galaxy. However, subsequent spectroscopic follow-up (see \S \ref{sec:specdetails}) showed that instead the transient, arc, and two sources were all at a common redshift of $z=0.1537\pm0.0002$ with velocity shifts of up to 150 km s$^{-1}$. %This is consistent with the 2dFGRS redshift of the central galaxy \citep[2dFGRS TGS493Z064, $z=0.1543$;][]{2001MNRAS.328.1039C}. 

\subsection{Photometry}

\subsubsection{DECam}

Following the initial epoch reported by \citet{2025GCN.42691....1H}, we continued monitoring the transient using DECam. In total, seven epochs of multifilter imaging was obtained between 2025-11-14 and 2025-12-31 (Table \ref{tab:lc}). Our DECam observations were analyzed using our GPU-enabled difference imaging pipeline (Figure \ref{fig:Discovery}) for rapid transient detection and photometry (Hu et al., in prep.), built on the high efficiency \texttt{SFFT} algorithm \citep{Hu2022}. Each science image was subtracted against archival DECam reference images using \texttt{SFFT} with aperture photometry calibrated to Legacy Survey imaging. A full account of the DECam transient pipeline is deferred to Hu et al., (in prep.). These methods have previously been outlined in \citet{Cabrera2024,Hu2025}. The log of photometry is tabulated in Table \ref{tab:lc}. The lightcurve is displayed in Figure \ref{fig:lightcurve}.

\subsection{Spectroscopic Follow-up}
\label{sec:specdetails}

\subsubsection{SALT}
%Ignore first spectrum only comment on the second one...

To explore the initial lensing hypothesis, we obtained longslit optical spectroscopic observations of the host galaxy and ring complex with the 10-m class Southern African Large Telescope \citep[SALT;][]{Buckley2006} at Sutherland Observatory in Sutherland, South Africa on 2025-11-18. %Observations were obtained through SALT Gravitational Wave Director's Discretionary Time. 
Longslit spectroscopy was acquired with the Robert Stobie Spectrograph \citep[RSS;][]{Burgh2003} using the PG0700 grating and a slit width $1.5\arcsec$. The slit was placed at a position angle (PA) of 90 deg East of North to cover both the ring feature and the two archival sources. The spectra covered the observer frame wavelength range 3592 \AA\ to 7479 \AA\ with resolution $R=735$ at the central wavelength of 5580 \AA. The data were reduced using the RSS Long-slit spectra processing and extraction app \texttt{rsslsspectra}\footnote{\url{https://astronomers.salt.ac.za/wp-content/uploads/sites/71/2024/09/rsslsspectra.pdf}}. Inspection of the 2D spectrum showed common emission and absorption lines between the ring and two sources, confirming they shared a common redshift ($z=0.1537\pm0.0002$). We extracted traces associated to the ring and both galaxies (hereafter denoted as G1 and G2). The slit placement is displayed in Figure \ref{fig:DeepImage} (as the blue rectangle) and the extracted 1D spectra are shown in Figure \ref{fig:SALT}. 

\begin{table}[]
    \centering
    \caption{DECam photometry of SN 2025adpq.}
    \begin{tabular}{lcccc}
    \textbf{MJD} & \textbf{Instrument} & \textbf{Filter} & \textbf{AB mag} \\
    \hline
60993.07466	& DECam &	$g$ &	$20.50\pm0.03$ \\
60994.08708	& DECam &	$g$ &	$20.67\pm0.03$ \\
60995.09111	& DECam &	$g$ &	$20.72\pm0.03$ \\
60997.08167	& DECam &	$r$ &	$20.49\pm0.03$ \\
60997.08281	& DECam &	$i$ &	$20.83\pm0.04$ \\
60999.08505	& DECam &	$g$ &	$21.09\pm0.04$ \\
60999.08573	& DECam &	$r$ &	$20.59\pm0.03$ \\
60999.08643	& DECam &	$z$ &	$21.30\pm0.15$ \\
60999.08712	& DECam &	$i$ &	$20.97\pm0.04$ \\
61014.12318	& DECam &	$g$ &	$22.54\pm0.25$ \\
61014.12389	& DECam &	$r$ &	$21.39\pm0.06$ \\
61014.12458	& DECam &	$z$ &	$21.44\pm0.17$ \\
61014.12529	& DECam &	$i$ &	$21.30\pm0.12$ \\
61040.09422	& DECam &	$i$ &	$22.20\pm0.25$ \\
61040.09634	& DECam &	$g$ &	$22.80\pm0.25$ \\
61040.09702	& DECam &	$r$ &	$22.50\pm0.15$ \\
61040.09768	& DECam & $z$	& $<21.5$  \\
\hline
    \end{tabular}
    \label{tab:lc}
\end{table}

\subsubsection{Gemini}

A classification spectrum of the transient was obtained using the Gemini Multi-Object Spectrograph (GMOS-S) mounted on the Gemini South telescope on 2025-11-24 starting at 02:49:51 UT, corresponding to 10 d after discovery. The spectrum was obtained using the B480 grating (covering 4060\AA\, to 8080\AA) with the 1\arcsec\, slit. The slit was placed at a PA of 129.28 deg to cover the transient and the ring (Figure \ref{fig:DeepImage}). The data were reduced using the \texttt{Dragons} software package \citep{Labrie2019,Labrie2023}. An inspection of the stacked 2D spectrum (Figure \ref{fig:Gem2D}) revealed emission lines at a consistent redshift ($z=0.1540\pm0.0002$; with a velocity spread of roughly 200 km s$^{-1}$) at the location of the transient and along the arc, extending for $\sim$8.5\arcsec\, (Figure \ref{fig:GemArc+Transient}). We extracted 1D spectra at the location of the transient and of the arc. The extracted 1D spectra of the transient and ring are shown in Figure \ref{fig:GemArc+Transient}.

\subsection{Archival Imaging}
\label{sec:archival}

%\url{https://vizier.cds.unistra.fr/viz-bin/VizieR?-source=II/383}

%https://www.eso.org/sci/publications/announcements/sciann17289.html
%https://www.eso.org/sci/publications/announcements/sciann17696.html %link to science portal is here

%https://alasky.cds.unistra.fr/KiDS/CDS_P_KiDS_DR5_color-gri/

\begin{figure*}
    \centering
    \includegraphics[width=1\columnwidth]{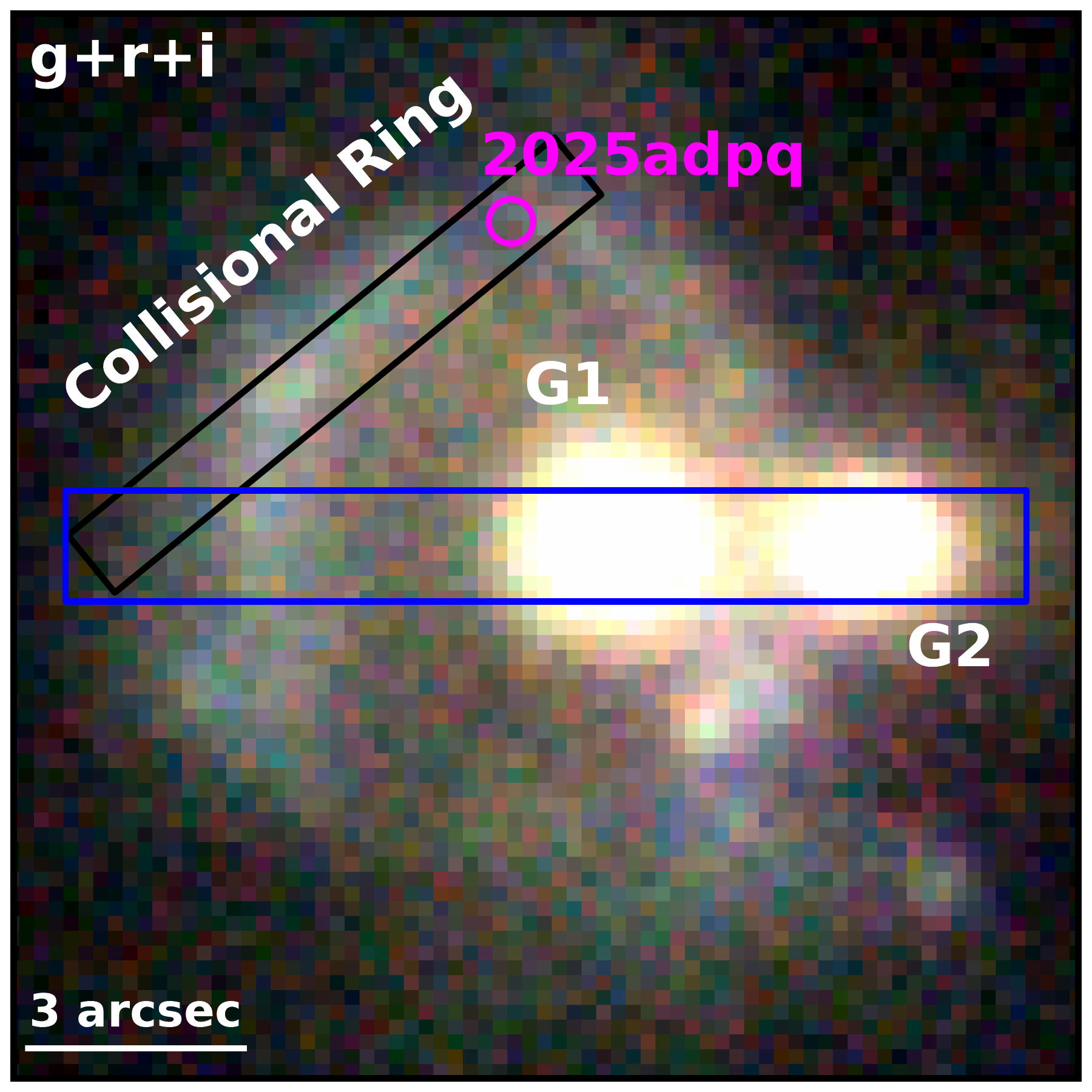}
    \hspace{-0.25cm}
    \includegraphics[width=1\columnwidth]{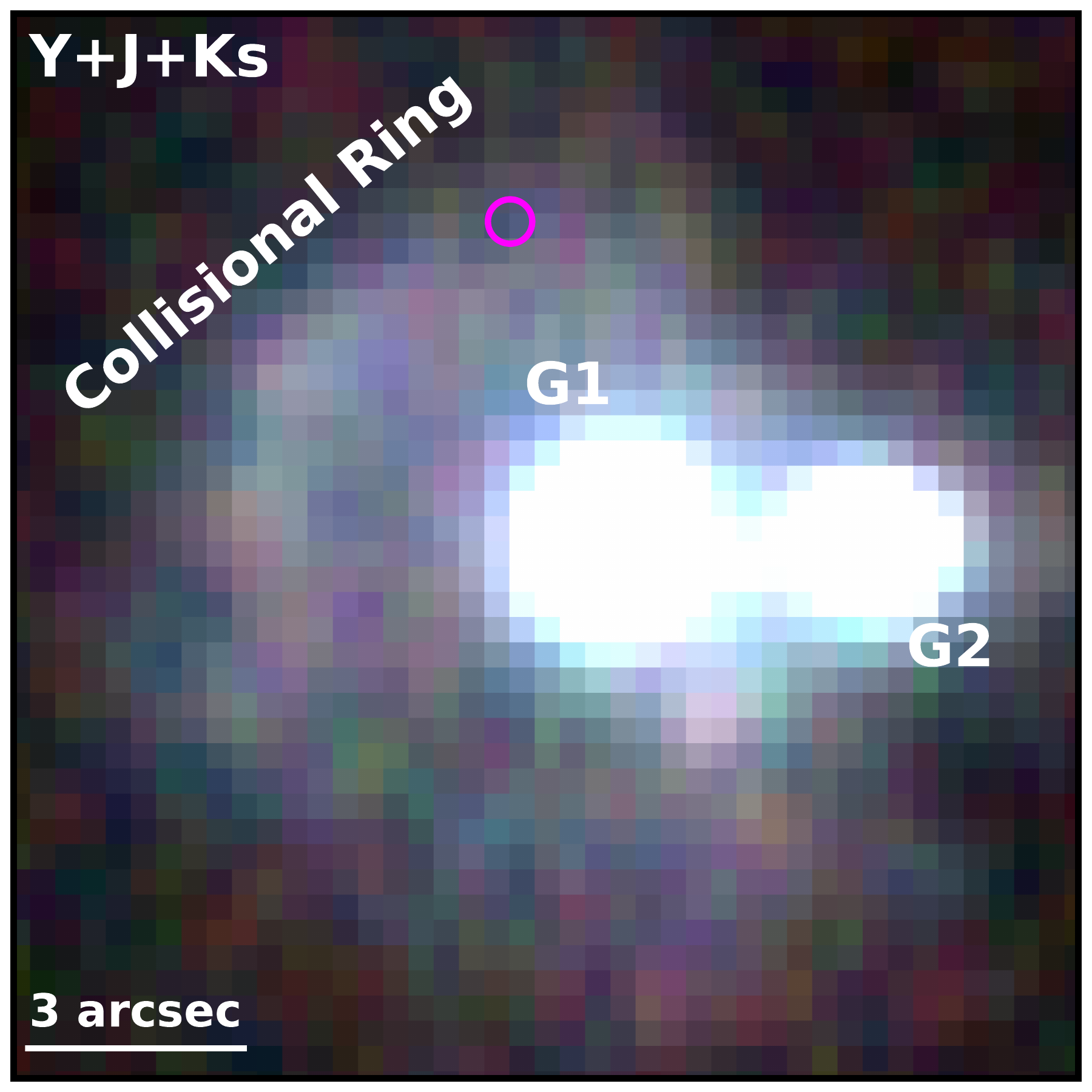}
    \caption{Left: Archival RGB image from the KiDS survey using the $gri$ bands. The transient position is shown by a magenta circle. The collisional ring and merging galaxies G1 and G2 are labeled. The black rectangle shows the placement of the Gemini GMOS-S slit and the blue rectangle shows the SALT RSS slit. Right: RGB image using VIKING near-infrared images in the $YJK_s$ bands. The scale and labels are the same. North is up and East is to the left.}
    \label{fig:DeepImage}
\end{figure*}

\begin{figure*}
    \centering
    \includegraphics[width=2\columnwidth]{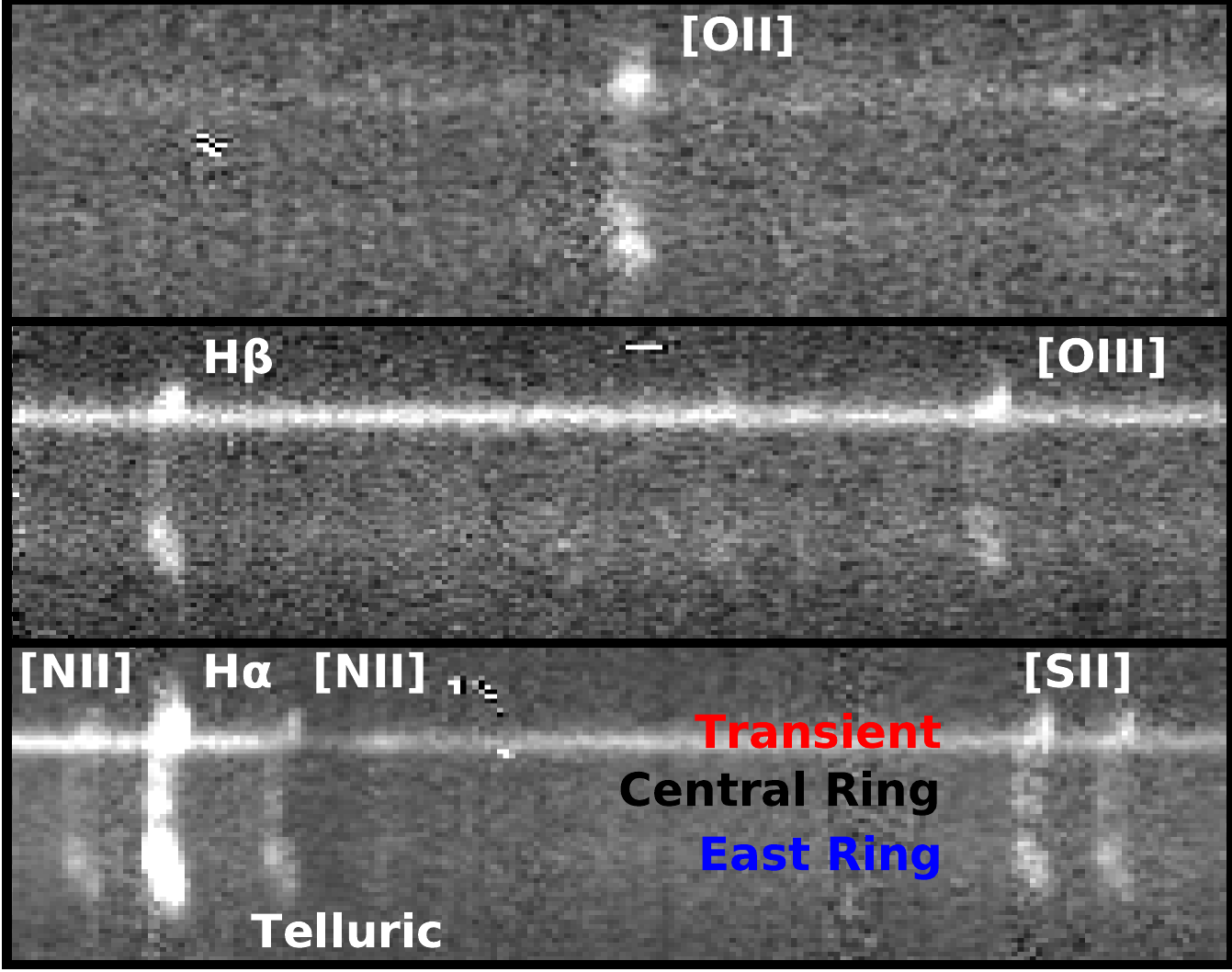}
    \caption{Visualization of the 2D Gemini GMOS-S spectrum obtained on 2025-11-24 ($\sim$\,10 days from discovery). Three different panels are shown that zoom in on relevant emission features. In the bottom panel the location of the transient, central arc, and eastern portion of the ring are labeled in red, black, and blue, respectively. The central region refers to a region of low surface brightness between the transient location and brightest portion of the ring as visible in Figure \ref{fig:DeepImage}. The spectral direction is the X-axis which goes from blue to red wavelengths from left to right. The spatial direction along the slit is the Y-axis. Each panel covers 250 \AA\, in width and is 8.5\arcsec\, in height. The scale is chosen to maximize the visibility of the emission lines. In the bottom panel the [NII] line is impacted by telluric absorption. A few cosmic rays can be seen throughout the figure, but they do not impact the emission features. 
    }
    \label{fig:Gem2D}
\end{figure*}

\begin{figure*}
    \centering
    \includegraphics[width=2\columnwidth]{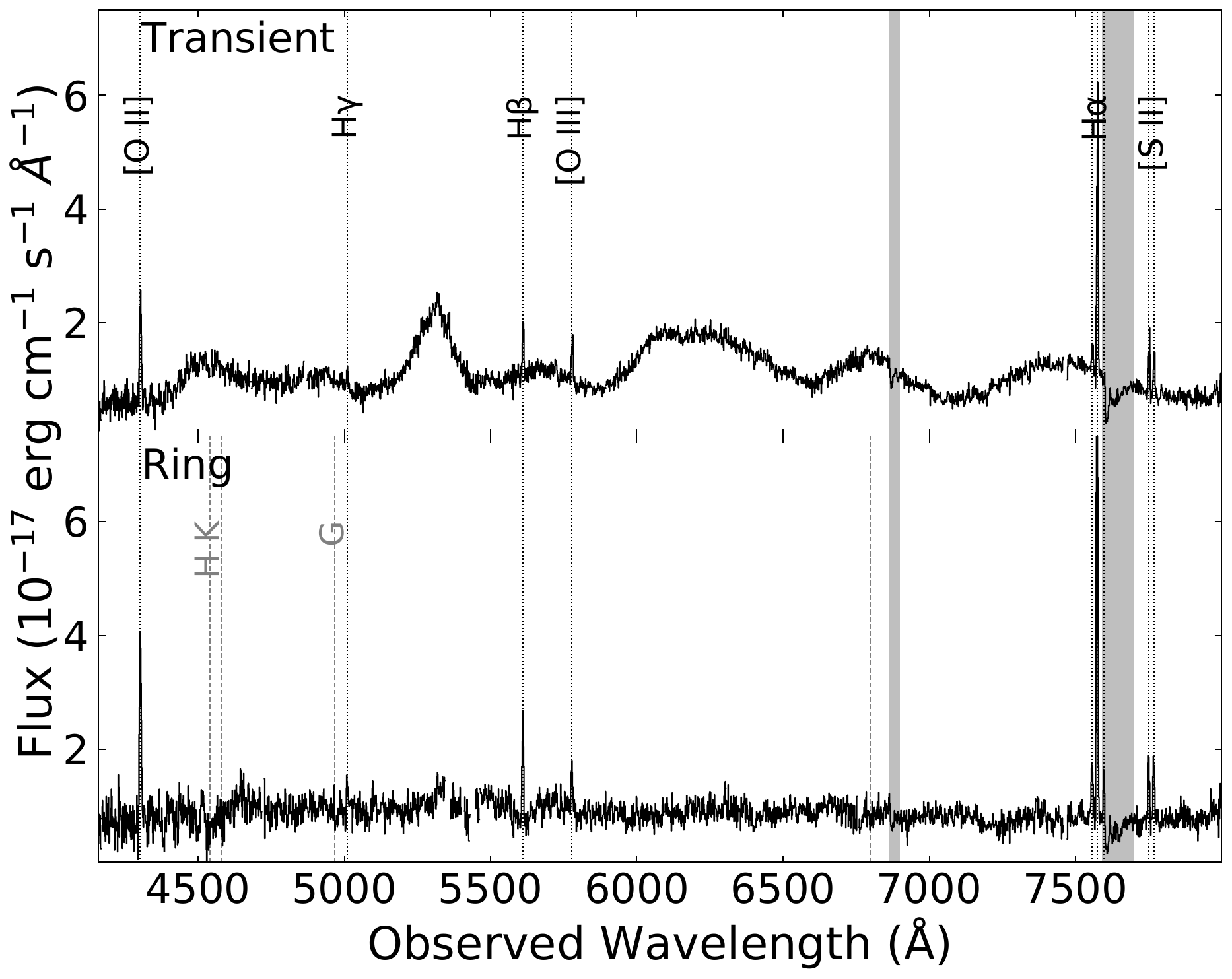}
    \caption{Gemini GMOS-S spectra of the transient and the eastward portion of the collisional ring. The slit placement is shown in Figure \ref{fig:DeepImage}. Numerous narrow emission and absorption features at a common redshift of $z$\,$=$\,$0.1540$ are labeled by the vertical lines. The gray shaded bands mark telluric regions.}
    \label{fig:GemArc+Transient}
\end{figure*}

While archival DECam imaging exists through the Legacy Survey, deeper optical imaging is available through the Kilo-Degree Survey (KiDS\footnote{\url{https://www.eso.org/sci/publications/announcements/sciann17696.html}}\footnote{\url{https://doi.eso.org/10.18727/archive/37}}; \citealt{2013Msngr.154...44D}) obtained using the OmegaCAM camera at the 2.6-m VLT
Survey Telescope and complementary near-infrared imaging obtained with VIRCAM on the 4.1-m VISTA telescope \citep{2006SPIE.6269E..0XD} as part of the VISTA Kilo-degree Infrared Galaxy Public Survey (VIKING\footnote{\url{https://www.eso.org/sci/publications/announcements/sciann17289.html}}\footnote{\url{https://doi.eso.org/10.18727/archive/59}}; \citealt{2013Msngr.154...32E}). We obtained the KiDS and VIKING optical and near-infrared photometry of the galaxies G1 and G2 through the ESO KiDS-DR5 multiband source catalog \footnote{\url{https://vizier.cds.unistra.fr/viz-bin/VizieR?-source=II/383}} \citep{Wright2024}. Forced multiband matched Gaussian Aperture and PSF (GAaP) photometry is available in the $ugriYJK_s$ filters. As the galaxies are relatively point-like and surrounded by contamination from either each other or the ring, we adopt this GAaP photometry. We additionally retrieved the available images in these filters from the ESO Science Archive Portal. The 5$\sigma$ limiting sensitivity in the $ugri$ filters is 24.2, 25.1, 25.0, and 23.7 AB mag, respectively, obtained under seeing $0.94\arcsec$, $1.18\arcsec$, $0.64\arcsec$, and $0.89\arcsec$, respectively. This imaging is shown in Figures \ref{fig:DeepImage} and \ref{fig:DeepImageAll1}. The $r$-band imaging from KiDS provides the best view of the collisional ring as it was obtain under the highest quality seeing.

%corrected for the negligible Galactic exinction $A_V=0.0285$ \citep{Schlafly2011}

%Extinction in each band...
%Au, Ag, Ar, Ai
%0.0454	0.0354	0.0245	0.0182
%AZ, AY, AJ, AH, AKs
%0.0135	0.0117	0.0076	0.0048	0.0032	

%5-sigma limiting sensitivity ugri = 24.2,25.1,25.0,23.7 under seeing 0.94, 1.18,0.64 0.89. Which is why the r-band looks the best, seeing is best.

%All AB mags
%G2
%u=21.2+/-0.04 [This seems wrong as no source is even seen in the u-band image...]
%g=19.53+/-0.004
%r=18.52+/-0.002
%i=18.11+/-0.005
%Y=17.62+/-0.003
%J=17.37+/-0.003
%H=17.18+/-0.005
%Ks=17.10+/-0.005

%G1
%u=20.84+/-0.03
%g=19.32 +/-0.003
%r=18.31 +/-0.0016
%i=17.91+/-0.004
%Y=17.41+/-0.002
%J=17.15+/-0.002
%H=16.97+/-0.004
%Ks=16.85+/-0.004

\begin{figure*}
    \centering
    \includegraphics[width=2\columnwidth]{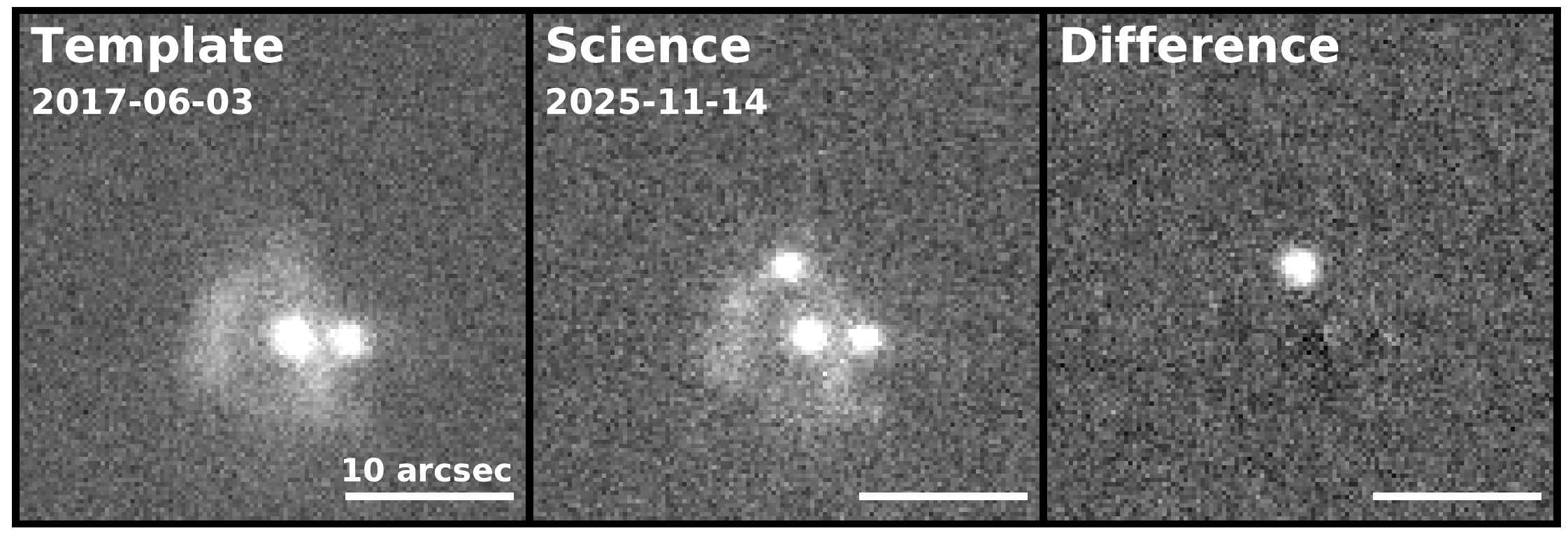}
    \caption{The DECam $g$-band discovery image of SN 2025adpq. The archival template is shown in the left panel. The middle panel shows the science image obtained on 2025-11-14 and the right panel is the difference image. North is up and East is to the left.}
    \label{fig:Discovery}
\end{figure*}

\begin{figure*}
    \centering
    \includegraphics[width=2\columnwidth]{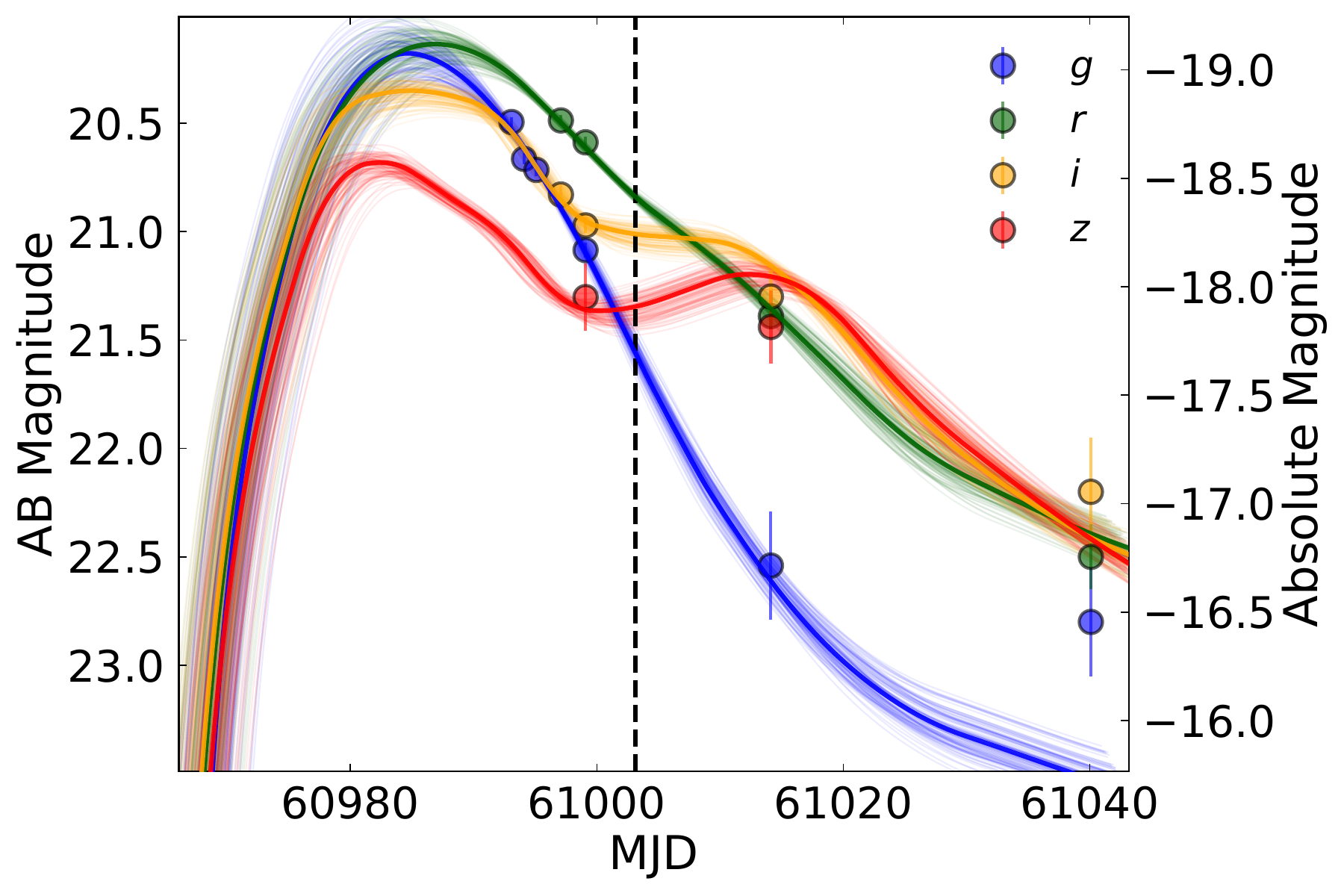}
    \caption{DECam optical ($griz$) lightcurve of SN 2025adpq in apparent (left side) and absolute (right side) magnitude. The best fit SALT2 model (thick lines) and 100 random models (thin lines) sampled from the posterior distributions are shown in each band (see \S \ref{sec:lcfit}). The black dashed line marks the epoch of our Gemini classification spectrum (see \S \ref{sec:class}).}
    \label{fig:lightcurve}
\end{figure*}

\begin{figure*}
    \centering
    \includegraphics[width=2\columnwidth]{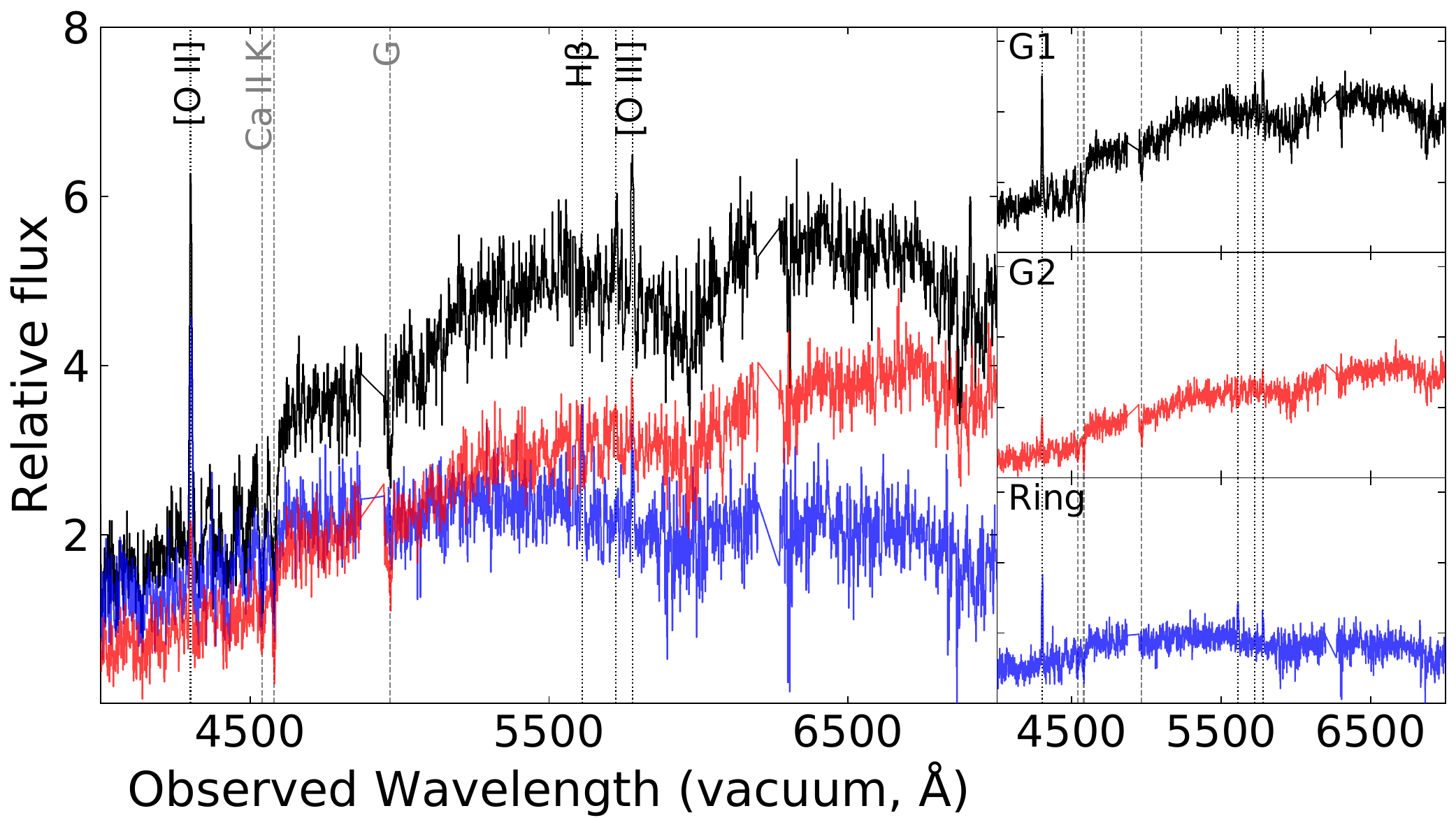}
    \caption{Optical spectra obtained with SALT/RSS of the galaxies G1 (black) and G2 (red) and the ring (blue). Narrow emission and absorption features at a common redshift $z$\,$=$\,$0.1537$  are shown by vertical lines. The right panels show the individual spectra for an easier comparison of the features. The chip gaps of SALT/RSS are clearly visible around $\sim$\,$4900\AA$ and $6200\AA$.}
    \label{fig:SALT}
\end{figure*}

\section{Results and Analysis}

\subsection{A Collisional Ring from a Major Galaxy Merger}

%Note lens models didn't like it even before we got a z... maybe an appendix as they have the figures already %https://repository.usfca.edu/cgi/viewcontent.cgi?article=1078&context=phys

%comment on if SN region looks SF or not in historic imaging - no obvious knot in imaging... BUT there is also an older stellar population in the arc signed by the H+K absorption... (though due to the transient emission we do not see H+K absorption at its location). 

%note the additional tail plausibly seen to the west of G2...

As visible in Figure \ref{fig:Discovery}, the transient, hereafter SN 2025adpq, lies on top of a ring-like feature. Figure \ref{fig:DeepImage} provides a clear pre-explosion view of the deep optical and near-infrared (OIR) imaging of the field. While this resembles a lensed\footnote{While an initial visual inspection showed this to be an interesting source complex for further investigation, prior to obtaining our definitive spectroscopic redshifts, our computed lens models were unable to adequately model this system using the methods of \citet{Huang2020lens}.} galaxy complex \citep[e.g.,][]{Huang2020lens}, optical spectroscopy obtained with SALT and Gemini reveals that the galaxies G1 and G2, the ring structure, and the transient, are all co-located at $z\approx0.1537$ to $0.1540$ (see \S \ref{sec:spec_analysis}). The spectroscopic redshift proved that this was instead a merging galaxy system with a collisional ring\footnote{Throughout the manuscript, we sometimes refer to this feature as the ``arc''.}. In this context, G2 can be referred to as the ``bullet''. % of stars ejected in the merger that still remains gravitationally bound. 
The luminosity ratio $(\Delta m_r\sim0.2)$ of G1 and G2, which is usually taken as a proxy for stellar mass ratio, is $L_2/L_1\sim0.8$, clearly falling in the major merger category \citep{Lambas2012}. We also note the presence of a possible tidal tail to the West of G2 (see Figure \ref{fig:ResidImage} for KiDS imaging); this feature is also visible in Legacy Survey imaging of the field.  

We note that at the transient location there appears to be a minimum between two arcs on the ring structure. While spectroscopy reveals strong star formation at the transient location (see \S \ref{sec:spec_analysis}), the strongest star formation occurs on the arc features visible in Figure \ref{fig:DeepImage}.

The projected offset between G1 and G2 is $\sim3.3\arcsec$, corresponding to $\sim9.2$ kpc. The transient is offset by $\sim4.1\arcsec$ from G1, corresponding to $\sim11.4$ kpc. We note that the ring is roughly $\sim$70 kpc in (projected) length (e.g., the circumference of a roughly $4\arcsec$ radius centered on G1).

\subsection{Spectral Analysis}
\label{sec:spec_analysis}

%https://docs.google.com/document/d/13WFPYCYc-wgiUbVO1FuQD2Vo4QXMSlQSFX_igyXD8Ng/edit?usp=sharing

We obtained optical spectra of the transient and host galaxy complex using the Gemini and SALT telescopes at two different position angles (see Figure \ref{fig:DeepImage}). The Gemini spectrum (black rectangle) covered the transient and ring to provide a classification spectrum and confirm the redshift of the transient and ring. The SALT spectrum (blue rectangle) measured the spectra of the two galaxies G1 and G2 and the ring at a slightly different position to Gemini. These spectra were both necessary to reject the lensing hypothesis and confirm the system is undergoing a major merger. 

Numerous narrow emission lines (e.g., [O\,\textsc{ii}], [O\,\textsc{iii}], H$\beta$) are detected from each of these sources and are clearly visible in the 1D and 2D spectra (Figure \ref{fig:Gem2D}). The emission lines are all at a common redshift of $z$\,$\approx$\,$0.1540$, confirming that the transient, ring, and two galaxies all occur at the same redshift and are linked to the same merging galaxy complex. As the SALT spectra cover only to $\sim6500\AA$ in the rest frame they do not cover H$\alpha$, only covering [O\,\textsc{ii}] through [O\,\textsc{iii}] (Figure \ref{fig:SALT}). Instead, the Gemini spectrum covers a broader wavelength range and detects [O\,\textsc{ii}] through H$\alpha$ (Figures \ref{fig:Gem2D} and \ref{fig:GemArc+Transient}).

The SALT spectra (Figure \ref{fig:SALT}) of G1, G2, and the ring reveal an old stellar population producing deep H, K, and G-band absorption, as well as ongoing star formation determined from bright [O\,\textsc{ii}] emission. Weak H$\beta$ emission is observed from G1 and the ring. There is a velocity shift of $\sim150$ km s$^{-1}$ between G1 and the ring (Figure \ref{fig:SALTzooms}). 

The Gemini spectrum of the transient and ring likewise reveal a variety of common lines as shown in Figure \ref{fig:Gem2D}. The transient spectrum allowed for the robust classification of the source as a Type Ia supernova as outlined in \S \ref{sec:class}. We find ongoing star formation at the transient location. This is determined from the bright nebular emission lines such as H$\alpha$, H$\beta$, [O\,\textsc{iii}] and [O\,\textsc{ii}] coinciding with the transient's trace in the 2D spectrum Figure \ref{fig:Gem2D}). However, a more active star forming region is identified at $<0.5\arcsec$ ($<$1.3 kpc) offset along the slit (Figure \ref{fig:Gem2D}), corresponding to the location of the upper ring region (Figure \ref{fig:DeepImage}). There is a clear velocity shift of $\sim$\,$120$ km s$^{-1}$ between the emission lines at these two locations along the slit (Figure \ref{fig:GemHalphaZoom}), suggesting the transient is independent of the more intense ongoing star formation in the upper ring (i.e., a chance alignment). This is supported by the detection of H and K band absorption (Figure \ref{fig:GemArc+Transient}) in the eastward ring (Figure \ref{fig:DeepImage}) that is indicative of an older population co-existing in the ring. We revisit this possibility later in the discussion. From the 2D spectrum (Figure \ref{fig:Gem2D}), we determine that the ongoing star formation extends across the ring for $8.5\arcsec$ along the slit, corresponding to a projected physical size of $\sim$24 kpc at $z$\,$\approx$\,$0.154$. This is consistent with the visual length of the ring covered by the slit (Figure \ref{fig:DeepImage}).

In particular, we focus on the Gemini spectra (Figure \ref{fig:GemArc+Transient}) which cover the transient location and the ring as a way to probe the local environment of the supernova. We use the penalized pixel-fitting (\texttt{pPXF}) software \citep{ppxf,ppxf2} to fit the optical spectra and derive the relevant emission line fluxes. We utilized the \texttt{E-MILES} stellar library \citep{EMILES}. We focus on the [O\,\textsc{ii}] doublet at $\lambda\lambda3726,3729$, nebular [O\,\textsc{iii}] at $\lambda\lambda4959,5007$, the Balmer lines H$\alpha$ and H$\beta$,  and [N\,\textsc{ii}] $\lambda6583$ adjacent to H$\alpha$. For the arc of the ring covered by the Gemini spectrum (see Figure \ref{fig:DeepImage}), we derive a near Case B recombination Balmer decrement H$\alpha/$H$\beta$\,$=$\,$2.65\pm0.13$, Baldwin, Phillips \& Terlevich (BPT; \citealt{BPT}) ratios of $\log\big([$O\,{\sc iii}$]/$H$\beta\big)$\,$=$\,$-0.26\pm0.05$ and $\log\big([$N\,{\sc ii}$]/$H$\alpha\big)$\,$=$\,$-0.80 \pm 0.05$ placing it firmly in the star forming region, a slightly subsolar gas phase metallicity 12 + log(O/H) $\sim 8.50\pm0.05$ ($\sim 0.6 Z_\odot$) derived from N2\footnote{We do note that the [$N\,{\sc ii}$] region is impacted by telluric absorption and inferences requiring its flux are likely less reliable.} and O3N2 \citep{PettiniPagel2004}. 
From the H$\alpha$ luminosity, we further derive a star formation rate (SFR) of $0.91\pm0.01 M_\odot$ yr$^{-1}$ \citep{Kennicutt1998} using a Chabrier \citep{Chabrier2003} initial mass function (IMF). This does not include the ``central'' region of the arc shown in Figure \ref{fig:Gem2D}, and just includes the bottom trace. 

%Gemini Arc Line Fluxes in W/m^2
%OII = 3.238e-18 +/- 1.3e-19
%Hbeta = 1.007e-18 +/- 4.7e-20
%OIII_5007 = 5.531e-19 +/- 5.6e-20
%Halpha = 2.67e-18 +/- 3.8e-20
%NII = 4.2e-19 +/- 5.1e-20

Due to the spatial and spectral resolution of GMOS and the close proximity of the additional star forming region to the transient within the slit, it is difficult to disentangle the emission line fluxes only underlying the transient's location (either physically or in projection). While there is clear H$\alpha$ emission coinciding with the transient's location in the slit, the H$\alpha$ flux from the more active region clearly extends to the transient's placement in the slit due either to its spatial extent behind the transient or the seeing limited spatial resolution of the observations, or both. The conclusion that this is unrelated to the transient is based on a few arguments. For starters, the [O\,\textsc{ii}], [O\,\textsc{iii}], and H$\beta$ emission show a  separation from the spatial location of the transient in the slit (Figure \ref{fig:Gem2D}), and this offset nebular emission appears brighter than the same emission lines at the location of the transient along the slit, which further supports the conclusion that the transient's location is less active. Secondly, the clear velocity shift (Figures \ref{fig:Gem2D} and \ref{fig:GemHalphaZoom}) with respect to the H$\alpha$ emission at the transient's trace  shows it is distinct both spatially (by $\sim$\,$1$ kpc) and in velocity space (by $\sim$\,$120$ km s$^{-1}$). The redshift of the more active region with respect to the transient's location implies it is likely behind the transient at a slightly further away region of the ring (the geometry of the ring and its likely inclination, with the eastward side closer to us, also support this). This is further supported by the inferred Balmer decrement of this active region (H$\alpha/$H$\beta$\,$=$\,$4.1\pm0.3$), which implies a high intrinsic extinction of $A_V$\,$\approx$\,$1.1\pm0.2$ mag. As the transient's spectrum is clearly not impacted by 1 magnitude of dust, when compared to typical SN Ia spectra (see \S \ref{sec:class}), this implies the dust lies behind the transient with respect to Earth. We note that the excess of intrinsic dust at this location could explain the observed gap between the east ring and upper ring (Figure \ref{fig:DeepImage}) that only coincides with the transient by chance alignment. 

For completeness, by extracting the line fluxes for the combined trace that includes the transient and the nearby actively star forming region (Figure \ref{fig:GemArc+Transient}), we derive a dust corrected SFR of $2.0\pm0.2M_\odot$ yr$^{-1}$ based on H$\alpha$ \citep{Kennicutt1998}. This is consistent with the dust corrected SFR of $2.3\pm0.5M_\odot$ yr$^{-1}$ derived using [O\,\textsc{ii}] \citep{Kewley2004}. We note that without the dust correction the SFR inferred from H$\alpha$ is $\sim$\,$0.85M_\odot$ yr$^{-1}$. Using the N2 and O3N2 diagnostics \citep{PettiniPagel2004}, we find a near solar metallicity of 12 + log(O/H) $\sim 8.65\pm0.05$.

\begin{figure*}
    \centering
    \includegraphics[width=1\columnwidth]{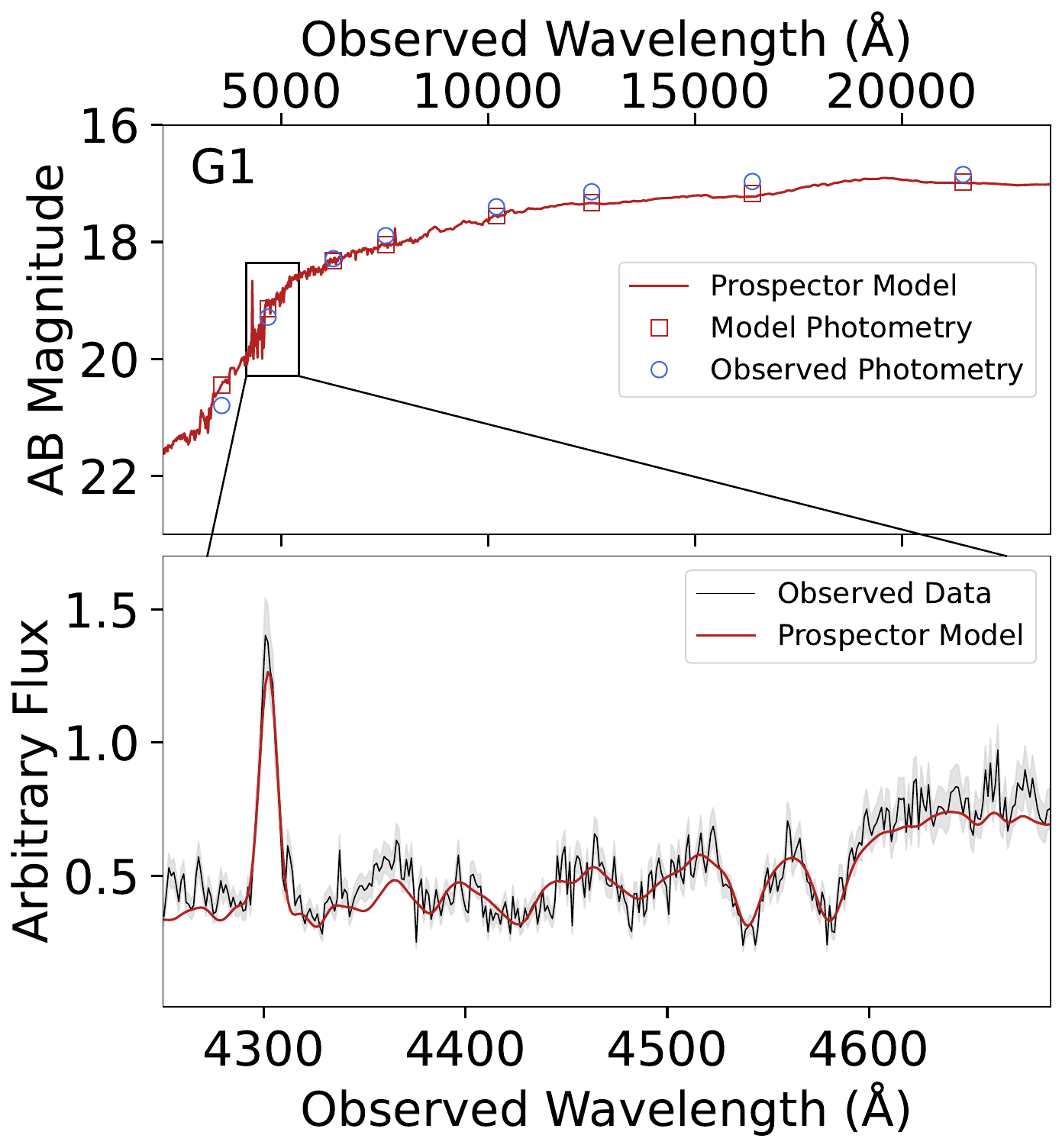}
    \includegraphics[width=1\columnwidth]{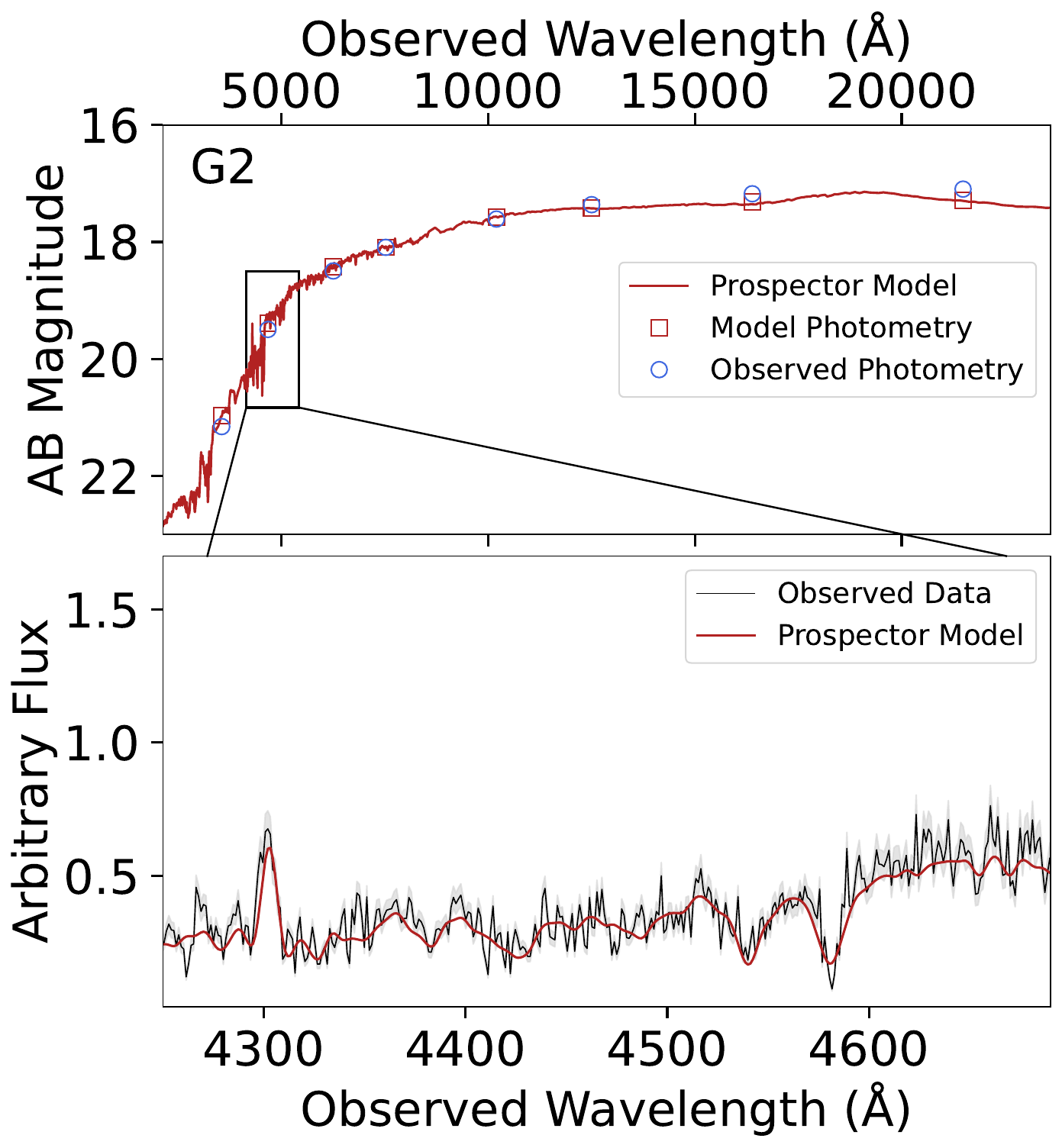}
    \caption{Results of our \texttt{prospector} modeling of the SEDs of G1 (left) and G2 (right). The top panel shows the best fit SED and model photometry (red squares) compared to the observed photometry (blue circles). The bottom panel shows the best fit spectrum (red) compared to the observed spectrum (black). We zoomed in on the most important spectral features, including the [O\,\textsc{ii}] doublet and Calcium H and K absorption lines. }
    \label{fig:prospectorSEDs}
\end{figure*}

\subsection{Galaxy SED Modeling}
\label{sec:prospector}

%non-parametric methods
%https://iopscience.iop.org/article/10.3847/1538-4357/ab133c/pdf
%https://iopscience.iop.org/article/10.3847/1538-4357/ad7e15/pdf
%https://arxiv.org/html/2509.12308v1
%https://iopscience.iop.org/article/10.3847/1538-4357/ab1d5a
%https://iopscience.iop.org/article/10.3847/1538-4357/aa5ffe

%16 parameters for G2
%(zsol,zgas,Ugas, dust1,dust2,mass, 5 logsfr ratios, sigmasmooth,specnorm)

In order to derive the galaxy properties, we performed spectral energy distribution (SED) modeling using \texttt{prospector} \citep{Leja2017,Johnson2019,Johnson2021}. We jointly fit the photometry and optical spectra, which were both corrected for the negligible Galactic extinction $A_V=0.0285$ mag \citep{Schlafly2011}. We adopt a \citet{Chabrier2003} initial mass function (IMF) and an intrinsic dust attenuation $A_V$ using a Milky Way extinction law \citep{Cardelli1989}, where we also account for additional dust around young stars. We adopt a nonparametric star formation history (SFH) as outlined in \citet{Leja2019nonpar,Leja2019nonpar2}. We use 7 bins to describe the SFH. The first two bins are between $0$\,$-$\,$30$ Myr and $30$\,$-$\,$100$ Myr to describe the young stellar population and ongoing star formation. We then apply three logarithmically spaced bins until $0.8t_\textrm{univ}$, where $t_\textrm{univ}$ is the age of the Universe (11.8 Gyr) at $z$\,$=$\,$0.154$. The final two bins are between $0.8t_\textrm{univ}$ to $0.9t_\textrm{univ}$ and $0.9t_\textrm{univ}$ to $t_\textrm{univ}$, which allows for a maximally old stellar population \citep{Park2024}. We utilize the continuity prior \citep{Leja2019nonpar} for the SFH with a Student-T prior ($\sigma$\,$=$\,$0.3$, $\nu$\,$=$\,$2.0$) for the change in the logarithm of the star formation rate ($\Delta \log$ SFR) between adjacent bins. We include nebular emission lines using the photoionization code \texttt{Cloudy} \citep{Ferland2013}. In order to better match the optical spectra, we also fit for the gas phase metallicity (decoupled from the stellar metallicity) and gas ionization parameter. We modeled the spectral continuum with a 12th order Chebyshev polynomial, a smoothing term, and a normalization term. We let the redshift $z$ vary by a small amount ($\pm0.005$) around the spectroscopic redshift of $0.154$ to account for slight velocity shifts. The synthetic SEDs derived from these model parameters were calculated using the flexible stellar population synthesis (\texttt{FSPS}) code \citep{Conroy2009} using a Planck cosmology \citep{Planck2020}. In total we fit the photometry and spectra using 16 free parameters. All fits were performed using \texttt{emcee} \citep{emcee} with 64 walkers for 4096 steps.

We modeled the SEDs of the merging galaxies G1 and G2 using KiDS and VIKING photometry covering the $ugriYJHK_s$ bands. We supplemented the photometry with the optical spectra obtained with SALT (Figure \ref{fig:SALT}). We find that the \texttt{prospector} models are capable of reproducing the observed photometry and spectra of G1 and G2. The best fit model SEDs are displayed in Figure \ref{fig:prospectorSEDs}, corner plots are shown in Figures \ref{fig:prospectorCorner1} and \ref{fig:prospectorCorner2}, and the star formation history is shown in Figure \ref{fig:SFH}. We find that the galaxies have a consistent stellar mass ($\log(M_*/M_\odot)$\,$\approx$\,$10.5$, with G2 potentially slightly more massive\footnote{We note that by using the KiDS GAaP (\S \ref{sec:archival}; \citealt{Wright2024}) we may be underestimating the true brightness of the sources, leading to slightly smaller stellar mass estimates.}. This confirms that the galaxies are undergoing a major merger, defined by having a mass ratio near unity. We find that both galaxies formed the bulk of their stars at early times (as likely required to match the H and K band absorption) and have had consistent, but lower, star formation in the last 500 Myr (see Figure \ref{fig:SFH}). We find G1 (SFR=$0.41^{+0.08}_{-0.05}M_\odot$ yr$^{-1}$) has a higher SFR than G2 (SFR=$0.10^{+0.02}_{-0.01}M_\odot$ yr$^{-1}$), which is consistent with presence of stronger [O\,\textsc{ii}] emission in the optical spectra (Figures \ref{fig:SALT} and \ref{fig:SALTzooms}). Both galaxies have slightly subsolar gas phase metallicities, with low gas ionization, and near solar stellar metallicity. Overall their properties are found to be quite similar.

\begin{figure*}
    \centering
    \includegraphics[width=2\columnwidth]{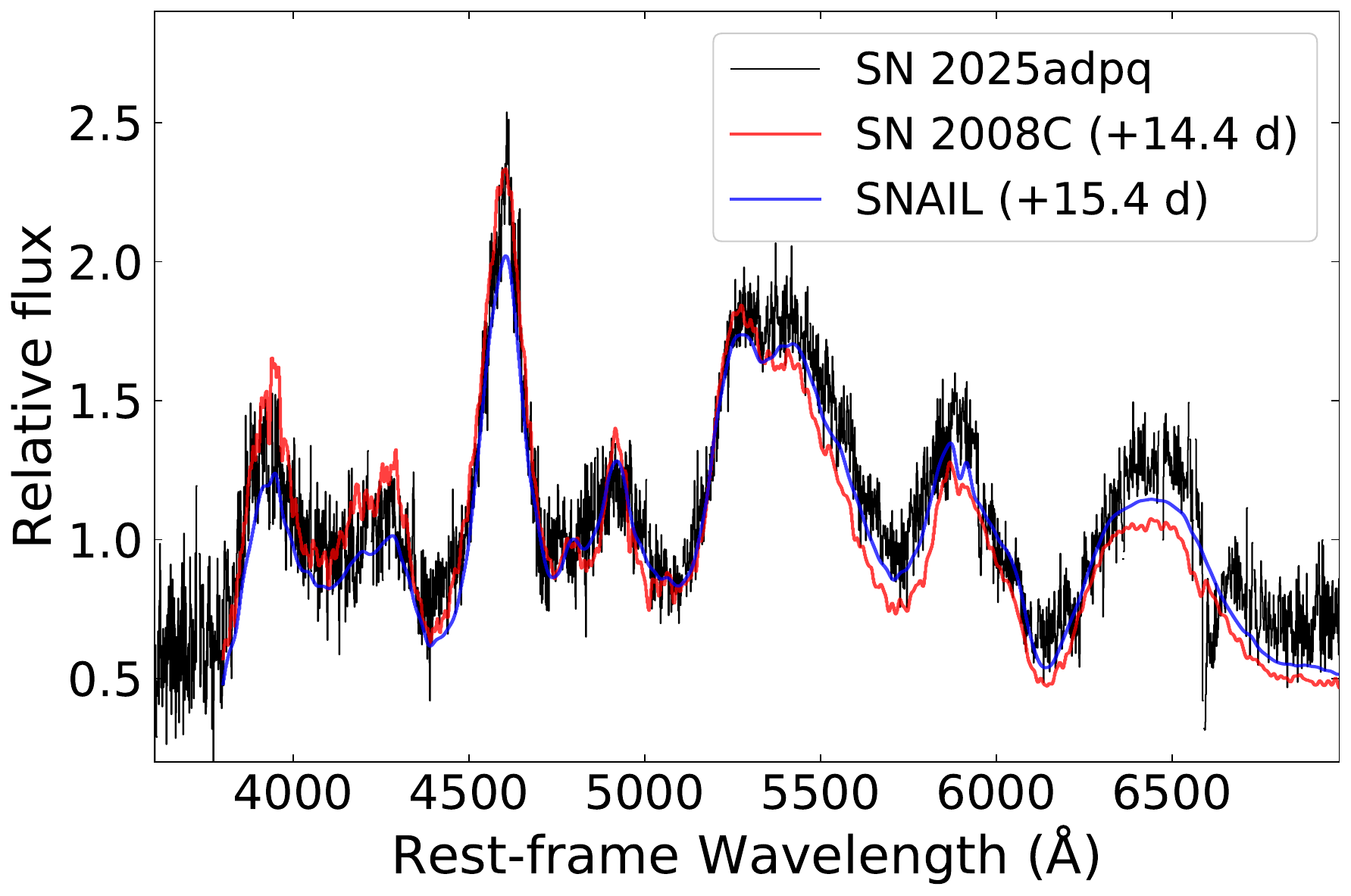}
    \caption{Classification spectrum (black) of SN 2025adpq obtained with Gemini GMOS-S. The best match SN Ia template spectra derived from \texttt{SNAIL} \citep{Hu2022SNAIL} are shown as blue a red lines. The red spectrum is the observed spectrum of SN 2008C \citep{2017AJ....154..211K} and the blue spectrum is a \texttt{SNAIL} interpolation. The spectrum has been clipped of the narrow emission lines shown in Figure \ref{fig:GemArc+Transient}. Telluric absorption is visible slightly above 6500 \AA. }
    \label{fig:GemTransient}
\end{figure*}

\subsection{Supernova Classification}
\label{sec:class}

%https://ui.adsabs.harvard.edu/abs/2022ApJ...930...70H/abstract
%https://iopscience.iop.org/article/10.3847/1538-4357/ac5c48

We performed a spectral classification of the Gemini spectrum of SN 2025adpq using Next Generation SuperFit (NGSF) and SuperNova Artificial Inference by Lstm neural networks (\texttt{SNAIL}\footnote{\url{https://github.com/thomasvrussell/snail}}; \citealt{Hu2022SNAIL}). The initial NGSF matches favored a SN Ia classification at $z$\,$=$\,$0.16$, consistent with the spectroscopic redshift of $0.1540$. The Ia classification led us to employ the \texttt{SNAIL} methodology to identify the best match from a Ia only template bank. \texttt{SNAIL} can interpolate the spectral sequence of the template bank and identify more accurate matches. We fixed the redshift to $z$\,$=$\,$0.1540$. \texttt{SNAIL} identified the best match template spectrum as SN 2008C, a normal SN Ia \citep{2017AJ....154..211K}, at 15.4 d post-peak. This match is shown in Figure \ref{fig:GemTransient}. The classification of the source as a SN Ia is further supported by the luminous discovery magnitude, which requires a peak luminosity in excess of $-19$ absolute magnitude (Figure \ref{fig:lightcurve}).

\subsection{Supernova Lightcurve Modeling}
\label{sec:lcfit}

We modeled the DECam multiband light curve (Figure \ref{fig:lightcurve}) of SN 2025adpq using the \texttt{SNCosmo}\footnote{\url{https://sncosmo.readthedocs.io/en/stable/}} \citep{SNCosmo} implementation of the SALT2 Type Ia supernova spectral time series model \citep{SALT2,Mosher2014}, fixing the redshift to $z=0.154$ and fitting for the standard SALT2 parameters $\{t_0, x_0, x_1, c\}$, where $t_0$ denotes the epoch of rest frame $B$-band maximum light in MJD, $x_0$ is the overall flux normalization, $x_1$ controls the lightcurve width (stretch), and $c$ encodes the color (including intrinsic color and reddening). The observed AB magnitudes in the $griz$ bands were converted to fluxes (with propagated uncertainties) and fit by minimizing $\chi^2$ to obtain a maximum likelihood solution. We then sampled the posterior probability distribution of the fitted parameters using Markov Chain Monte Carlo (via \texttt{emcee}; \citealt{emcee}), adopting uniform bounds on $\{t_0, x_0, x_1, c\}$. We ran the fit using 60 walkers for 4096 steps. We verified that the chain length  was greater than 50 autocorrelation times for each parameter to ensure convergence. The best fit model is shown in Figure \ref{fig:lightcurve} and a corner plot is displayed in Figure \ref{fig:lcfitcorner} in Appendix \ref{sec:lcfitcorner}.

We obtain the best fit values of $t_0 = 60985.5\pm1.1$ d, $x_1=-0.82\pm0.31$, and $c = 0.07\pm0.03$ with $\chi^2/$dof$= 18.6/12$. These values are in agreement with standard SALT2 fits from larger samples \citep[e.g,][]{Nicolas2021,Rigault2025}. The derived peak time is about a week ($7.5\pm1.1$ d) before discovery, and the classification spectrum is about 2 weeks after peak ($17.6\pm1.1$ d). This is consistent with our expectations from the spectral analysis (Figure \ref{fig:GemTransient}), which provided a best match at $\sim$\,15.4 d post-peak to the Type Ia SN 2008C \citep{Hu2022SNAIL}.

%=== SALT2 best fit (z fixed at 0.15400) ===
% t0 = 60985.47685087169
% x0 = 0.00012633973111233037
% x1 = -0.8249879625329921 
% c = 0.07384409407329112
% chisq/ndof = 18.60 / 12

\section{Discussion}

\subsection{Properties of the Merger and Collisional Ring}
\label{sec:ringdisc}
%Other ring refs
%https://adsabs.harvard.edu/full/1976ApJ...209..382L
%https://arxiv.org/abs/astro-ph/0703600
%https://adsabs.harvard.edu/full/1995ApJ...453..641W
%https://arxiv.org/abs/2211.05785
%https://ui.adsabs.harvard.edu/abs/1987ApJ...318..103A/abstract
%https://ui.adsabs.harvard.edu/abs/1996FCPh...16..111A/abstract
%https://ned.ipac.caltech.edu/level5/Sept01/Appleton/frames.html
%https://ui.adsabs.harvard.edu/abs/1997AJ....113..201A/abstract
%https://ui.adsabs.harvard.edu/abs/1996ApJ...468..532A/abstract
%https://academic.oup.com/mnras/article/417/2/835/983327
%https://arxiv.org/abs/astro-ph/0703600

%\begin{figure}
%    \centering
%\includegraphics[width=\columnwidth]{figs/project-test.png}
%    \caption{Orientation etc... Magenta = 3.4x3.78 arcsec with angle 10 deg; green = 3.79x4.58 arcsec with angle 25 deg}
%    \label{fig:placeholder}
%\end{figure}

Collisional ring galaxies are necessarily rare due to their transient nature and short timescale ($\lesssim$ 500 Myr; e.g., \citealt{Wong2006,Renaud2018,Elagali2018eagle}). The majority of known rings are identified in the local Universe, with $\sim$\,$130$ known at $z$\,$\sim$\,$0.002$\,$-$\,$0.09$ \citep{Madore2009}. Additional examples at higher redshifts are necessary to expand the sample, and Pika's Halo provides a clean example at $z$\,$=$\,$0.15$, though others are known at both similar \citep[e.g,][]{Conn2011,Smith2012} and higher \citep[e.g,][]{vanDokkum2025ring,Khoram2025ring} redshifts. 

While the Vera C. Rubin Observatory \citep[e.g.,][]{Ivezic2019} will enable visual identification of a larger sample, spectroscopy is required to distinguish between an Einstein ring and collisional ring or tidal tail. In addition, at higher redshifts the ring would have a smaller angular size, and its discovery will be a seeing limited quantity, where the ring can blend with the larger host galaxy, adding difficulty in its recognition. For example at $z\approx0.5$, a $\sim$\,10 kpc radius corresponds to $\sim$\,$1.5\arcsec$, but the same feature would be $\sim$\,1/3 of the brightness compared to its value at $z=0.15$ due to the surface brightness dimming by $(1+z)^{-4}$. These changes in its visual size and brightness could lead the ring to potentially blend with the galaxy depending on its morphology and angular size, and the inclination of the system with respect to our line-of-sight. However, at $z\approx0.25$, such features would be easier to confirm, which still provides a larger volume available for Rubin. In any case, Pika's Halo adds to the growing sample of spectroscopically confirmed collisional rings beyond the very local Universe \citep[e.g.,][]{Madore2009,Conn2011,Smith2012,Pasha2025,vanDokkum2025ring,Khoram2025ring}.

The nearly equal luminosities and stellar masses of G1 and G2 indicate that the system has a mass ratio of order unity and is clearly in the process of undergoing a major merger. The collisional ring shows that G1 and G2 recently underwent a head-on bullseye collision in which the bullet (G2) collided with the disk of the SN's parent galaxy (G1), producing a collisionally driven density wave that both compresses the interstellar medium and dynamically rearranges pre-existing stars \citep[e.g.,][]{StruckMarcellLotan1990,HernquistWeil1993,AppletonStruckMarcell1996}. The optical spectra support this picture by showing widespread nebular emission associated with ongoing star formation in the ring, while Calcium H and K absorption demonstrates that an old stellar population also contributes to the ring light.

We can use the ring geometry to place basic constraints on the inclination angle and timescale. Under the assumption that the intrinsic shape of the ring is a perfect circle, we can infer the inclination with respect to our line-of-sight. The approximation of a circular ring structure is based on  theoretical simulations 
of head-on collisions driving nearly symmetric radial density waves \citep[see, e.g.,][]{LyndsToomre1976, AppletonStruckMarcell1996} and commonly 
adopted in observational deprojections of collisional rings \citep[e.g.,][]{Amram1998, Fogarty2011, Yuan2020ring,Khoram2025ring}. 
The ring has a projected shape (see Figure \ref{fig:DeepImageAll1}) of an ellipse with semi-major axis $a$\,$\approx$\,$3.8\arcsec$ (10.5 kpc) and semi-minor axis $b$\,$\approx$\,$3.4\arcsec$ (9.4 kpc). As the major axis (10.5 kpc) is unaffected by inclination, we can determine the inclination as $b/a$\,$=$\,$\cos i$, which yields $i$\,$\approx$\,$26.5$ deg. 

Assuming a constant expansion velocity of 100 km s$^{-1}$ requires that the collision occurred within the last $\approx$\,$100$ Myr to produce the observed size. Based on observations of other collisional rings reported in the literature \citep[e.g.,][]{Higdon1996,Higdon1997,Amram1998,Fogarty2011}, we consider a range of expansion velocities of $50$\,$-$\,$150$ km s$^{-1}$, which would yield a ring age between 70 to 200 Myr. This is a reasonable age for a collisional ring and is comfortably within the short ring lifetime inferred for rings produced in simulations \citep[e.g.,][]{Wong2006,Renaud2018,Elagali2018eagle}. This timescale is also broadly consistent with the current projected separation of G1 and G2. As the line between G1 and G2 lies roughly along the semi-minor axis of the ring, the deprojected separation is $(9.2\,\textrm{kpc})/\cos i$\,$\sim$\,$10.2$ kpc, which is similar to the ring's radius of $\sim$\,$10.5$ kpc. Thus, the bullet would require a similar relative velocity with respect to G1 of  order $90$\,$-$\,$100$ km s$^{-1}$ over $\sim$\,$100$ Myr. The relative velocity of G1 and G2, as well as the ring's expansion speed, are variable (decreasing) with time \citep[e.g,][]{Martinez-Delgado2023}, and while the true three dimensional trajectory of this system is uncertain, the simple consistency (in our back of the envelope calculation) between the ring age and the present separation of G1 and G2 supports a recent passage as the formation channel for the ring. 

Our optical spectra further reveal coherent velocity structure along the ring, with line centroid shifts that change systematically with position along the arc.
As shown in Figures \ref{fig:Gem2D} and \ref{fig:GemHalphaZoom}, there is a clear line-of-sight velocity shift along the arc with a roughly $\sim$135 km s$^{-1}$ blueshift from the active star formation at the top of the ring to the transient position. The velocity along the upper East (referred to as the Central ring in Figure \ref{fig:Gem2D}) facing arc of ring is then roughly constant until the lower edge of the ring probe by the GMOS-S slit, where then the velocity experiences a redshift of $\sim$150 km s$^{-1}$. 
These changes in velocity occur along changes in direction (or position angle; PA) of the arc of the ring. The maximum change in velocity identified along this arc of the ring (spanning roughly 70 deg of PA around the ring) is $\sim$\,$350$ km s$^{-1}$, with sharper changes coincident with bends in the projected arc. 

In an expanding ring, the observed line of sight velocity is set by the combination of radial expansion, any residual disk rotation, and projection effects from inclination, so gradients of this magnitude are plausible even for modest physical expansion speeds when the slit samples multiple azimuths around the ring \citep[see, e.g.,][]{Khoram2025ring}. A spatially resolved velocity field from integral field spectroscopy would be the most direct way to separate expansion from rotation and to provide a stronger test of whether the ring kinematics align with a single density wave launched by the passage of G2 in a bullseye collision with G1.

\subsection{Comparison to SN Ia Environments}

Type Ia supernovae occur across a broad range of host galaxies and environments \citep{Sullivan2006,Lampeitl2010,Pan2014,Senzel2025,Ruppin2025}, but their rate is closely tied to stellar mass and to a wide delay time distribution that extends from hundreds of Myr to several Gyr \citep{Totani2008,Maoz2012,Maoz2012rev}. The identification of SN 2025adpq's host galaxy (G1) as a massive elliptical is consistent with the broader population. Instead, its peculiarity is that the supernova lies in a star forming structure at an offset of 11.4 kpc from G1. We have identified the structure as a dynamically produced collisional ring where both old and young populations coexist.
The location of SN 2025adpq along the ring suggests a simple interpretation that is consistent with the delay time expectations for Type Ia progenitors. The collisionally driven pressure wave both triggered star formation and swept out pre-existing disk stars into the expanding ring \citep[e.g.,][]{HernquistWeil1993,AppletonStruckMarcell1996,Renaud2018}. The Calcium H and K absorption observed in the ring demonstrates that an old population of stars formed pre-collision are present along the structure. Moreover, the inferred lightcurve properties of the supernova (\S \ref{sec:lcfit}), in particular the low stretch $x_1$\,$\approx$\,$-0.8$, have been previously linked to older systems \citep[see][]{Nicolas2021}.

The collision of G2 with G1 therefore provides a natural mechanism to relocate long-lived Type Ia progenitor systems away from their parent galaxies, allowing a Type Ia explosion to occur at large projected offsets without requiring an intrinsically young (short delay time) progenitor channel. The interpretation of SN 2025adpq as an old system stripped from its host galaxy is also supported by the bulk of star formation in G1 occurring $\sim$\,$1$ Gyr ago (see Figure \ref{fig:SFH}), consistent with an old delayed progenitor channel \citep{Totani2008,Maoz2012,Maoz2012rev}. However, we cannot conclusively rule out a rapid formation channel due to recent collisionally induced star formation. A young progenitor would  be consistent with inferences that a significant fraction of SN Ia progenitors occur through a prompt formation channel with delay times $<$\,$0.5$ Gyr \citep{Childress2014,Nicolas2021,Wiseman2022} which is in good agreement with the expected age of the collisional ring (\S \ref{sec:ringdisc}), see also \citet{Wong2006,Fogarty2011,Renaud2018}. Deep spatially resolved spectroscopy to constrain the stellar continuum in the ring, together with improved constraints on the ring age, would enable a more direct comparison between the inferred progenitor delay time and the merger timescale.

The environment of SN 2025adpq is also relevant to the broader class of apparently hostless Type Ia supernovae and large offset events \citep[e.g,][]{GalYam2003,Sand2011,Graham2015,Qin2024,Strolger2025,SN2021hem}. A number of Type Ia supernovae have been reported with no obvious host in wide field imaging, including cases in galaxy clusters where intracluster stars can be produced through tidal stripping \citep{GalYam2003,Sand2011,Graham2015}. More recent work has emphasized that many apparently hostless Type Ia events in the field \citep[e.g,][]{SN2021hem} may instead be associated with faint or ultra diffuse dwarf hosts \citep{Qin2024,Strolger2025}, as opposed to hypervelocity stars \citep[e.g.,][]{Zinn2011,SN2021hem}. These hostless supernovae are generally identified based on their extremely large offsets from all candidate host galaxies, with dimensionless directional light radii $d_\textrm{DLR}$\,$>$\,$4$\,$-$\,$5$ \citep[see, e.g.,][]{Strolger2025}. In the case of SN 2025adpq, considering the offset from G1, we derive $d_\textrm{DLR}$\,$\approx$\,$3$, which, even neglecting the ring, is small enough to be considered a robust host association in general Ia samples \citep{Sullivan2006,Gupta2016,Sako2018}.

While SN 2025adpq does not represent a truly hostless SN Ia, the location of SN 2025adpq presents a complementary pathway to produce an offset supernova where the local surface brightness of the parent galaxy is low. This is the case regardless of whether SN 2025adpq formed through an old progenitor system that relocated to the outskirts of its parent galaxy or if it was formed through a faster channel in a recent burst of star formation produced by the expanding density wave. At higher redshifts, the decrease in observed surface brightness could render the ring undetectable, in which case the supernova could be misclassified as a hostless (or large offset) event despite being physically connected to a major merger. However, collisional rings are an unlikely explanation for extreme (100 kpc) offset events such as the hostless SN 2021hem \citep{SN2021hem}, and instead provide a reasonable explanation for SN located off their host galaxy's light at a few to tens of kpc. More broadly, the discovery of SN 2025adpq in Pika's Halo motivates targeted searches for faint tidal and collisional structures around large offset Type Ia supernovae, particularly in samples where host association is ambiguous \citep[e.g.,][]{Strolger2025}.

We note that SN 2025adpq lies at a larger redshift than the typical horizon for SN Ia detection of current pre-LSST wide field optical surveys (e.g., the Zwicky Transient Facility; ZTF). For example, while the ZTF SN Ia Data Release 2 (DR2) contains 3628 nearby ($z < 0.3$) Ia supernova \citep{Rigault2025ztf-dr2-overview} the median redshift is only $z\approx0.065$ with only 11\% of the sample at $z>0.1$ and $1\%$ above $z>0.15$, where SN 2025adpq was discovered. The onset of Rubin will greatly expand this horizon of detected SN Ia for photometric classification \citep[e.g.,][]{Gris2023,Petrecca2024}, but the large scale spectroscopic classification of SN Ia at $z>0.1$ will be challenging due to their faint peak magnitudes ($\gtrsim20.5$ AB mag).

\begin{table*}
\centering
\caption{ 
Matched TNS transients within a projected radius of 30 kpc around the \citet{Madore2009} local collisional ring galaxy sample and the samples of \citet{Timmis2017,Shamir2020}. The column ``Ring?'' gives our visual classification of whether the transient lies on the ring structure. A cross ($\times$) in the final column marks a sibling system, i.e., a CRG with more than one matched TNS transient.}
\label{tab:madore_tns_30kpc}
\small
\setlength{\tabcolsep}{4pt}
\resizebox{\textwidth}{!}{%
\begin{tabular}{lllcccccc}
\hline
CRG & TNS Object & Type & $z_{\rm CRG}$ & $z_{\rm TNS}$ & Sep. ($^{\prime\prime}$) & Sep. (kpc) & Ring? & Sibling \\
\hline
AM 0545-434 & SN 2023abv  & SN Ia  & 0.0578 & 0.057 & 32.5 & 22.6 & No  &  \\
AM 1006-380 & AT 2018gfd  & \dots  & 0.016488 & \dots    & 60.6 & 20.3 & No  &  \\
AM 1354-250 & AT 2024rlc  & \dots  & 0.020514 & \dots    & 5.6  & 2.3  & No  & $\times$ \\
AM 1354-250 & AT 2025dhr  & \dots  & 0.020514 & \dots    & 24.3 & 10.1 & Yes & $\times$ \\
AM 2308-324 & SN 2025khv  & SN Ia  & 0.037783 & 0.037 & 14.3 & 10.8 & Yes &  \\
AM 2353-291 & SN 2019fkq  & SN Ia  & 0.029791 & 0.045$^a$ & 7.8  & 4.6  & No &  \\
Arp 010     & AT 2025abev & \dots  & 0.030381 & \dots    & 27.3 & 16.6 & Yes &  \\
Arp 143     & AT 2017oe   & \dots  & 0.0142 & \dots    & 3.3  & 1.0  & No  & $\times$ \\
Arp 143     & SN 2016bam  & SN II  & 0.0142 & 0.014 & 37.0 & 10.7 & Yes & $\times$ \\
Arp 147     & SN 2018gwr  & SN II  & 0.032209 & 0.032 & 10.0 & 6.4  & No  &  \\
Arp 284     & SN 2023pso  & SN Ib  & 0.009333 & 0.00933 & 2.0  & 0.4  & No  &  \\
IC 1908     & AT 1992bv   & \dots  & 0.027466 & \dots    & 7.4  & 4.1  & No  &  \\
NGC 4774    & AT 2023acdq & \dots  & 0.027929 & \dots    & 10.4 & 5.8  & Yes & $\times$ \\
NGC 4774    & AT 2017ivd  & \dots  & 0.027929 & \dots    & 14.2 & 8.0  & Yes & $\times$ \\
NGC 4774    & SN 2021cjd  & SN IIP & 0.027929 & 0.027929 & 2.2  & 1.3  & Yes & $\times$ \\
\hline
SDSS-Ring-12 & AT 2021hlt & \dots & \dots & \dots & 4.9 & 3.2 & Yes & \\
PS1-Ring-36  & SN 2021wxl & SN II & 0.0473 & 0.0473 & 4.1 & 2.6 & Yes & \\
PS1-Ring-39  & SN 2016ejf & SN Ib/c & \dots & 0.0280 & 3.7 & 2.4 & Yes & \\
PS1-Ring-124 & SN 2019cec & SN II & 0.0252 & 0.02597 & 13.5 & 8.7 & Yes & \\
PS1-Ring-155 & AT 2022cux & \dots & 0.0279 & \dots & 0.4 & 0.3 & No & $\times$ \\
PS1-Ring-155 & SN 2019svd & SN Ib/c & 0.0279 & 0.02796 & 13.6 & 8.8 & Yes & $\times$ \\
\hline
\end{tabular}%
}
\vspace{1mm}

{\footnotesize
%$^{a}$ No catalog recession velocity is reported for AM 0545$-$434 in the \citet{Madore2009} tables. We therefore adopted the median redshift of the measured-redshift CRG sample, $z_{\rm med}=0.035$, to define the 30 kpc search radius only; it should not be interpreted as a measured host-galaxy redshift. \\ %Turns out AM 0545$-$434 has a redshift in HyperLeda.
$a$ The redshift of the SN is likely slightly off as it clearly lies in this galaxy complex.
}
\end{table*}

\begin{table*}
\centering
\caption{
Sibling CRG systems, defined here as CRGs hosting more than one matched TNS transient within the 30 kpc search radius. }
\label{tab:madore_tns_siblings}
%\small
\setlength{\tabcolsep}{5pt}
\begin{tabular}{lccc}
\hline
CRG & $N_{\rm match}$ & Ring / Non-ring & Matched TNS Objects \\
\hline
AM 1354-250 & 2 & 1 / 1 & AT 2025dhr, AT 2024rlc \\
Arp 143     & 2 & 1 / 1 & SN 2016bam, AT 2017oe \\
NGC 4774    & 3 & 3 / 0 & SN 2021cjd, AT 2023acdq, AT 2017ivd \\
PS1-Ring-155 & 2 & 1 / 1 & SN 2019svd, AT 2022cux \\
\hline
\end{tabular}
\end{table*}

\subsection{A Search for Other Supernovae in Collisional Ring Galaxies}

%\textcolor{red}{supernova siblings ref https://arxiv.org/pdf/2112.14819 and https://iopscience.iop.org/article/10.3847/1538-4357/adad60 etc?}
%\citep{AndersonSoto2013,Scolnic2020,Graham2022,Salo2025}

%[clarify which AT and which SN]
%Not in a ring
%2018gfd - Gaia only; not in ring according to LS DR11 (private imaging)
%2023abv
%1992bv
%2023pso
%2018gwr 
%2017oe
%2024rlc

%[clarify which AT and which SN]
%In ring
%2023acdq - NGC 4774
%2017ivd - NGC 4774 
%2021cjd - NGC 4774
%2016bam
%2025abev - DESI spec??
%2019fkq
%2025khv
%2025dhr

%NGC 4774 is a SN factory!

%ARP 143 is also a supernova sibling - one in central galaxy one in ring...
%AM 1354-250 is also a supernova sibling - one in central galaxy one in ring...

In order to understand the rarity of SN 2025adpq, we crossmatched the \citet{Madore2009} catalog\footnote{\url{http://cdsarc.cds.unistra.fr/viz-bin/cat/J/ApJS/181/572}} of collisional ring galaxies (CRGs) against the public Transient Name Server (TNS) catalog\footnote{\url{https://www.wis-tns.org/}}, adopting a fiducial projected search radius of 30 kpc. We retained only the ring nucleus entries in the \citet{Madore2009} component tables (i.e., objects labeled ``:RN''), and excluded the listed companions. For CRGs with cataloged recession velocities, we converted the projected search radius to an angular radius through
\begin{equation}
\theta_{30} = \frac{30\, {\rm kpc}}{D_A(z_{\rm CRG})}\times 206265^{\prime\prime},
\end{equation}
where $D_A$ is the angular diameter distance for a flat $\Lambda$CDM cosmology with $H_0=70$ km s$^{-1}$ Mpc$^{-1}$ and $\Omega_m=0.3$. The \citet{Madore2009} compilation contains 132 ring nuclei in total; 89 have cataloged recession velocities, while 43 do not. For those 43 systems, we adopted the median redshift of the CRG sample, $z_{\rm med}=0.035$, solely to define the angular search radius. At this typical redshift, 30 kpc corresponds to $\sim$\,$43\arcsec$; across the full CRG sample the adopted 30 kpc search radii span $18\arcsec$\,$-$\,$310\arcsec$.

We searched the full public TNS catalog (as of March 1, 2026) rather than restricting the sample to spectroscopically classified supernovae. This choice was deliberate because TNS contains not only post-2016 discoveries but also a heterogeneous set of pre-2016 transients from legacy surveys. We therefore imposed no discovery date cut. The resulting search returned 15 TNS transients within 30 kpc of 11 CRGs (8\% of the sample). These comprise seven classified supernovae and eight unclassified transients.

We visually inspected archival imaging of each matched transient to identify its position relative to the ring morphology of its parent CRG and classified it as either \emph{in the ring} or \emph{not in the ring}. Under this classification, 7 of the 15 matched transients are located in the ring and 8 are not. Those 8 events not located in the expanding ring are associated to either the parent CRG galaxy or the ``bullet'' and are likely still physically associated to the CRG system. Three CRG systems host multiple matched transients and therefore qualify as sibling systems \citep[e.g.,][]{AndersonSoto2013,Scolnic2020,Graham2022,Salo2025}. NGC 4774 is the most prolific object in the sample (a ``supernova factory''), with three matched transients, all visually coincident with the ring. Two additional CRGs, Arp 143 and AM 1354-250, each host a mixed sibling configuration in which one transient lies in the ring and one lies interior to the ring/central body. We note that none of these CRGs hosting multiple transients has two associated classified supernovae, and it is a mix match between classified and unclassified events and thus not true ``supernova siblings''. However, it is statistically likely that those unclassified transients are in fact supernovae.

Table \ref{tab:madore_tns_30kpc} lists the full matched sample, including the CRG and TNS identifiers, transient type, CRG and TNS redshifts where available, angular and projected separations, the visual ring classification, and a sibling flag. Table \ref{tab:madore_tns_siblings} summarizes the three multiple transient CRG systems. In this sample, the visually ring associated transients are AT 2025dhr, SN 2025khv,  AT 2025abev, SN 2016bam, AT 2023acdq, AT 2017ivd, and SN 2021cjd. The non-ring subsample consists of SN 2023abv, AT 2018gfd, AT 2024rlc, AT 2017oe, SN 2018gwr, SN 2019fkq, SN 2023pso, and AT 1992bv. It must be noted that some of these AT transients were discovered by private surveys (e.g., PS1) and we could not further vet their lightcurves and classifications, nor whether they are difference imaging artifacts. 
%We also note that one matched transient in the 30 kpc sample, SN 2023abv in AM 0545$-$434, is associated with a CRG lacking a catalog recession velocity; for that system the median CRG redshift was used only to set the search radius and should not be interpreted as a measured host redshift; the redshift is measured from the SN classification spectrum and available on TNS. [This galaxy has a hyperleda redshift that agrees with the SN so we removed this statement and used that redshift.]

We note that the \citet{Madore2009} catalog is comprised of sources located at very low redshifts ($z$\,$<$\,$0.09$) where SN completeness is higher, and the prospects of missing SN detections should be lower. However, the catalog is based on the Arp–Madore Southern Peculiar Galaxies Catalog \citep{ArpMadore1987} and lies majority in the Southern hemisphere where so far transient surveys have been more limited. However, as 110 out of 132 CRGs (83.3\%) in the catalog lie in the Vera C. Rubin Observatory's planned survey footprint, there is hope to expand the sample of known transients in these galaxies in the coming decade. 

The lack of a complete all sky collisional ring galaxy catalog is a significant roadblock to accurate statistics of their supernova and transient rates. As an attempt to correct the Southern bias of the \citet{Madore2009} catalog, we also performed the same search over a sample of an additional 49 collisional rings identified in SDSS \citep{Shamir2020} and PS1 \citep{Timmis2017} imaging. We note that other catalogs exist for non-collisional ring galaxies such as polar ring galaxies \citep[e.g.,][]{Schweizer1983,Whitmore1990} or resonance rings, but we do not consider these other classes of much more common rings in our searches. There are comprehensive DESI Legacy Imaging searches \citep{Zhang2025ringsLS} using machine learning, but they do not separate between ring classes and our visual inspection revealed that the purity of the sample is limited with many identified ``rings'' belonging to normal looking spiral galaxies; we therefore did not consider this sample in our analysis either. In any case, even with the addition of these 49 Northern hemisphere (SDSS and PS1) rings, our sample is far from complete and there is no uniformly selected, vetted sample of collisional ring galaxies. We note that even some \citet{Madore2009} selected collisional rings were not identified in the algorithmic based analyses \citep[e.g., Table 3 of][]{Timmis2017}, strongly suggesting that the completeness of these samples is very likely low.

These catalogs \citep{Timmis2017,Shamir2020} span similar redshifts to the \cite{Madore2009} sample with ranges of $z$\,$<$\,$0.08$ and roughly 35\% fall within the Rubin LSST footprint. By performing the same crossmatching analysis on these additional 49 collisional rings, we identify 6 new matches transients, with 5 located within the ring and 1 located in the primary galaxy. Among the 5 in-ring transients, 4 are classified supernovae (SN 2016ejf, SN 2019cec, SN 2019svd, and SN 2021wxl). We also identify PS1-Ring-155 as a likely supernova sibling as it hosts both SN 2019svd and AT 2022cux. These host galaxy for added events are contributed by \citet{TimmisShamir2017} (5 objects) and \citet{Shamir2020} (1 object). We note that many of these galaxies lacked spectroscopic redshifts  and we adopted the median $z$ of the full sample to compute the 30 kpc matching radius. The information on these transients is shown in Tables \ref{tab:madore_tns_30kpc} and \ref{tab:madore_tns_siblings}.

To our knowledge, based on this analysis, we find that SN 2025adpq is only the 8th classified supernova identified in the ring of a collisional ring galaxy complex, and only the 10th transient in general. An exact determination is difficult due to the lack of complete all sky catalogs of collisional rings. We identified also two additional SN Ia located in a CRG complex, with SN 2025khv also lying in the expanding ring. Thus, SN 2025adpq is the second classified SN Ia located in a collisional ring (though 2025khv remains unpublished). Additionally, the unclassified transient AT 2024rlc has a peak absolute magnitude of $\sim$\,$-19$ according to its ATLAS lightcurve, and is likely a SN Ia.

\section{Conclusions}
\label{sec:conclusion}

We present the discovery of a Type Ia supernova, SN 2025adpq, in a collisional ring (dubbed Pika's Halo) formed by a major galaxy merger at $z=0.1540$. The supernova is offset by 11.4 kpc from the central galaxy, G1, and lies along the ring, which has a projected circumference of $\sim70$ kpc. While there is evidence for merger induced ongoing star formation in close proximity to the transient location, there is also evidence for an old stellar population (based on H and K absorption lines) in the ring. Based on this, we propose that the stellar progenitor is potentially an old system stripped from the host galaxy G1 by the pressure wave induced by the collision with the bullet galaxy G2. The pressure wave formed the observed collisional ring of star formation and the old progenitor system eventually exploded in the ring at a large offset. This interpretation is supported by the fact the bulk of star formation in the system occurred over 1 Gyr ago. However, we note that a rapid progenitor channel cannot be ruled out and that SN Ia progenitors have a broad range of delay times with evidence for both a ``tardy'' and ``prompt'' formation channel \citep{Childress2014,Wiseman2022}.

We performed an archival search of transients within known low redshift ($z$\,$<$\,$0.09$) collisional ring galaxies using the \citet{Madore2009} catalog. To our knowledge, based on this analysis, we find that SN 2025adpq is only the 8th classified supernova identified in the ring of a collisional ring galaxy complex, and only the 10th transient in general. An exact determination is difficult due to the lack of complete all sky catalogs of collisional rings. We identified also two additional SN Ia located in a collisional ring complex, with one also located within the expanding ring (SN 2025khv). An additional candidate SN Ia (based on its peak absolute magnitude) is found within a parent galaxy of another ring system (AT 2024rlc). We further identify 3 collisional ring galaxies that host candidate supernova siblings with multiple transients detected in those source complexes. As such, collisional ring galaxy complexes are clearly factories for the production of supernova of multiple types.

The location of SN 2025adpq (and SN 2025khv) in a collisional structure has implications for the offsets of Type Ia supernova from their host galaxies, especially observationally hostless Type Ia supernova at large offsets (ten to tens of kpc) with no visible underlying stellar population. However, the rarity of collisional rings, comprising only $\sim$\,$1\%$ of all ring galaxies,  versus all possible host galaxies suggests they cannot be a dominant factor in shaping the offset distribution of Type Ia supernovae.
In any case, the Vera C. Rubin Observatory Legacy Survey of Space and Time \citep{Ivezic2019} will aid in identifying low surface brightness tidal features around galaxies that have undergone recent major or minor mergers that may host such old stellar populations at large offsets, and can further reveal supernovae (Type Ia or CCSN) occurring in tidal tails and collisional rings. A full study of supernova rates in collisional rings could hold strong constraints on their formation and delay times as the age of the ring can in principle be dynamically derived.

\section*{ACKNOWLEDGEMENTS}

The authors thank the anonymous referee for thoughtful
feedback on the text, and some insightful suggestions that strengthened the manuscript. B. O. gratefully acknowledges support from the McWilliams Postdoctoral Fellowship in the McWilliams Center for Cosmology and Astrophysics at Carnegie Mellon University. 

This work used resources on the Vera Cluster at the Pittsburgh Supercomputing Center (PSC). We thank the PSC staff for helping set up our software on the Vera Cluster. The Vera Cluster is a dedicated computing resource for the McWilliams Center for Cosmology and Astrophysics at Carnegie Mellon University. 

This project used data obtained with the Dark Energy Camera (DECam), which was constructed by the Dark Energy Survey (DES) collaboration.
Funding for the DES Projects has been provided by the US Department of Energy, the US National Science Foundation, the Ministry of Science and Education of Spain, the Science and Technology Facilities Council of the United Kingdom, the Higher Education Funding Council for England, the National Center for Supercomputing Applications at the University of Illinois at Urbana-Champaign, the Kavli Institute for Cosmological Physics at the University of Chicago, Center for Cosmology and Astro-Particle Physics at the Ohio State University, the Mitchell Institute for Fundamental Physics and Astronomy at Texas A\&M University, Financiadora de Estudos e Projetos, Fundação Carlos Chagas Filho de Amparo à Pesquisa do Estado do Rio de Janeiro, Conselho Nacional de Desenvolvimento Científico e Tecnológico and the Ministério da Ciência, Tecnologia e Inovação, the Deutsche Forschungsgemeinschaft and the Collaborating Institutions in the Dark Energy Survey.

The Collaborating Institutions are Argonne National Laboratory, the University of California at Santa Cruz, the University of Cambridge, Centro de Investigaciones En\`ergeticas, Medioambientales y Tecnol\`ogicas–Madrid, the University of Chicago, University College London, the DES-Brazil Consortium, the University of Edinburgh, the Eidgenössische Technische Hochschule (ETH) Zürich, Fermi National Accelerator Laboratory, the University of Illinois at Urbana-Champaign, the Institut de Ci\'encies de l’Espai (IEEC/CSIC), the Institut de F\'isica d’Altes Energies, Lawrence Berkeley National Laboratory, the Ludwig-Maximilians Universit\:at M\:unchen and the associated Excellence Cluster Universe, the University of Michigan, NSF’s NOIRLab, the University of Nottingham, the Ohio State University, the OzDES Membership Consortium, the University of Pennsylvania, the University of Portsmouth, SLAC National Accelerator Laboratory, Stanford University, the University of Sussex, and Texas A\&M University.

Based on observations at Cerro Tololo Inter-American Observatory, NSF’s NOIRLab (NOIRLab Prop. ID 2023B-851374, PI: Andreoni \& Palmese), which is managed by the Association of Universities for Research in Astronomy (AURA) under a cooperative agreement with the National Science Foundation.
We thank Kathy Vivas, Alfredo Zenteno, and CTIO staff for their support with DECam observations. Some of the observations reported in this paper were obtained with the Southern African Large Telescope (SALT).

Based on observations obtained at the international Gemini Observatory, a program of NSF's OIR Lab, which is managed by the Association of Universities for Research in Astronomy (AURA) under a cooperative agreement with the National Science Foundation on behalf of the Gemini Observatory partnership: the National Science Foundation (United States), National Research Council (Canada), Agencia Nacional de Investigaci\'{o}n y Desarrollo (Chile), Ministerio de Ciencia, Tecnolog\'{i}a e Innovaci\'{o}n (Argentina), Minist\'{e}rio da Ci\^{e}ncia, Tecnologia, Inova\c{c}\~{o}es e Comunica\c{c}\~{o}es (Brazil), and Korea Astronomy and Space Science Institute (Republic of Korea). 

Based on data obtained from the ESO Science Archive Facility with DOI:
https://doi.org/10.18727/archive/37, and https://doi.eso.org/10.18727/archive/59 and on
data products produced by the KiDS consortium. The KiDS production team
acknowledges support from: Deutsche Forschungsgemeinschaft, ERC, NOVA and
NWO-M grants; Target; the University of Padova, and the University Federico II
(Naples). This publication has made use of data from the VIKING survey from VISTA at the ESO
Paranal Observatory, programme ID 179.A-2004. Data processing has been contributed by the
VISTA Data Flow System at CASU, Cambridge and WFAU, Edinburgh.

This research has made use of the NASA/IPAC Extragalactic Database (NED), which is funded by the National Aeronautics and Space Administration and operated by the California Institute of Technology.

This research has made use of the VizieR catalogue access tool, CDS, Strasbourg Astronomical Observatory, France (DOI : 10.26093/cds/vizier).

\appendix 
\section{Additional Imaging Figures}

\begin{figure*}
    \centering
  \includegraphics[width=1\columnwidth]{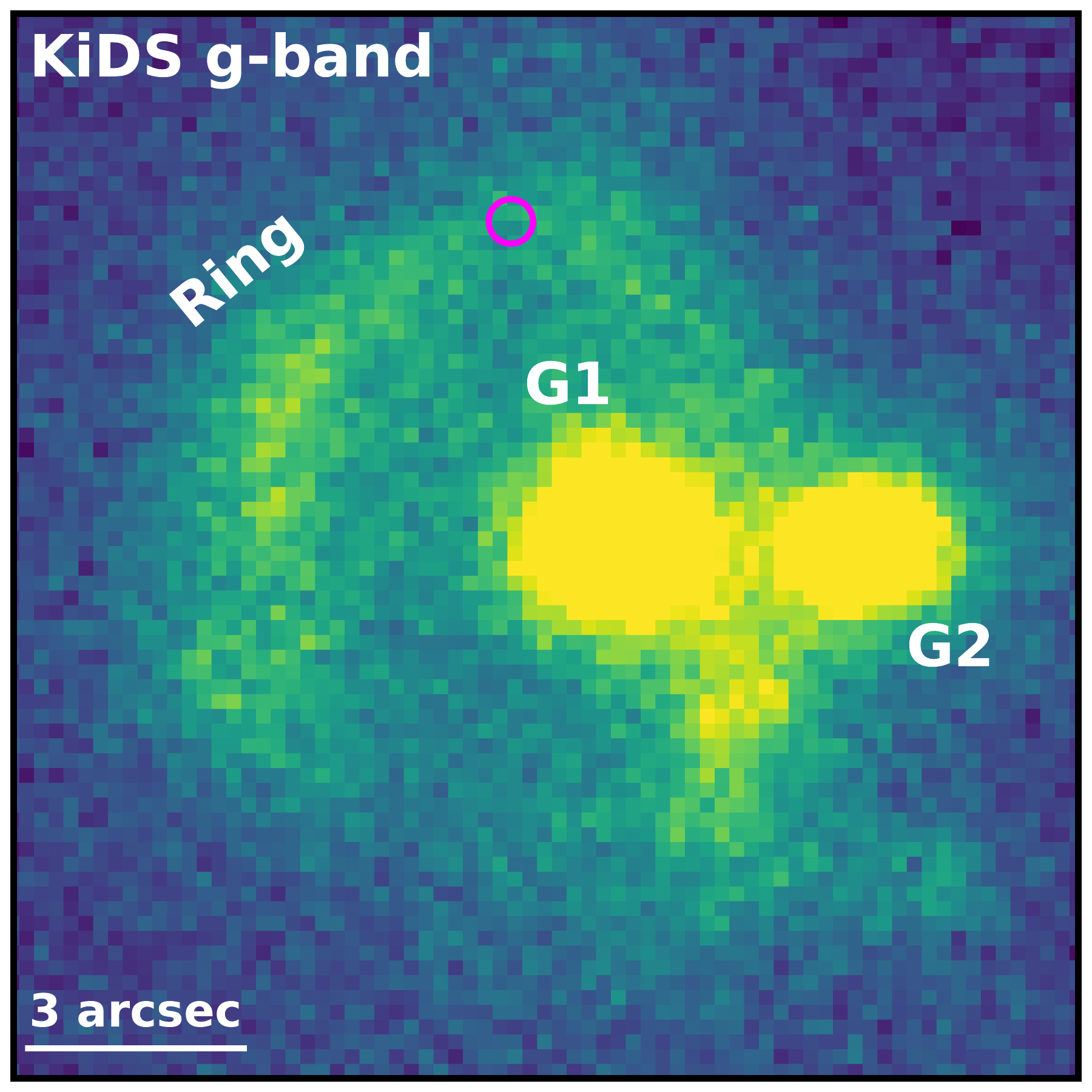}
    \hspace{-0.25cm}
        \vspace{-0.25cm}
    \includegraphics[width=1\columnwidth]{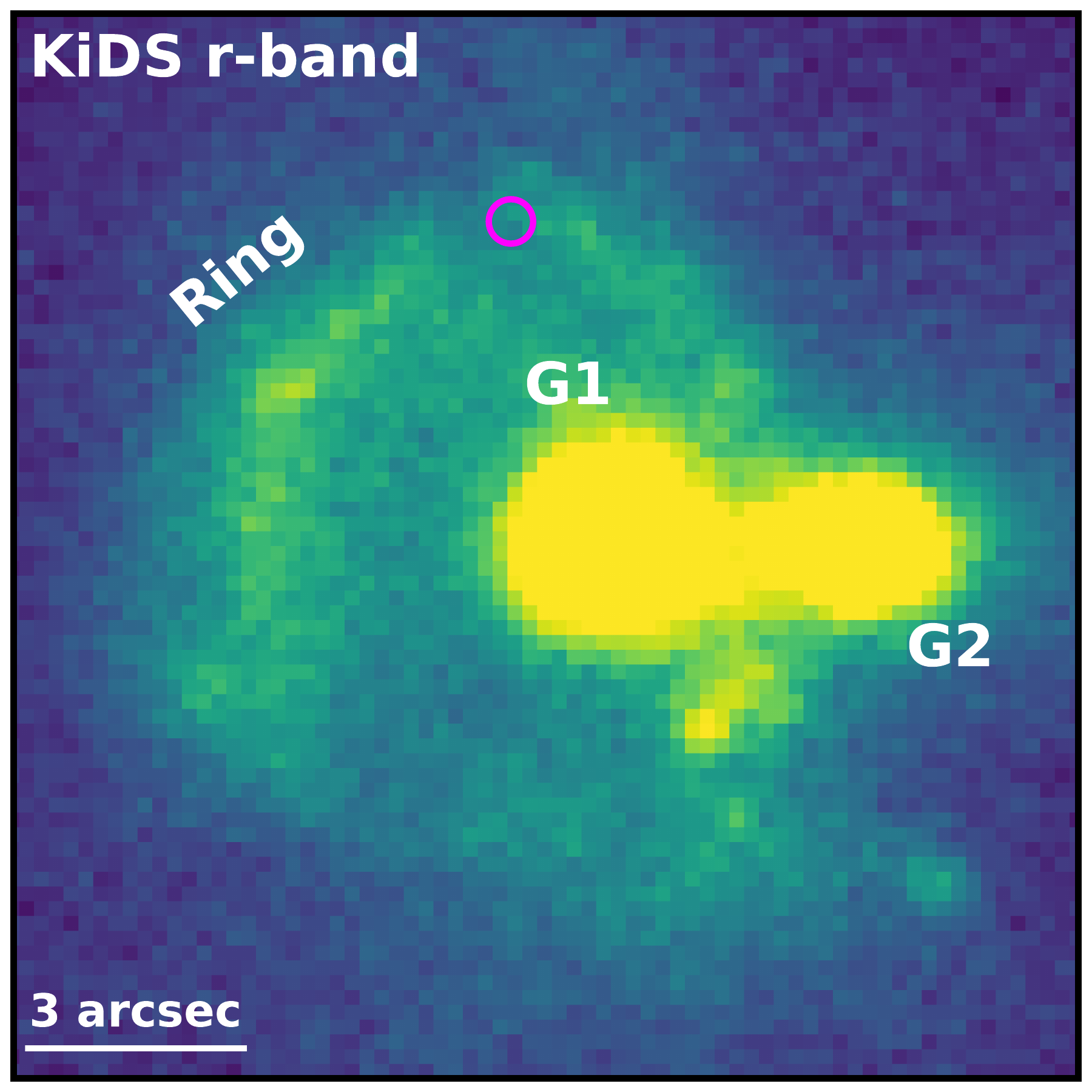}
    \includegraphics[width=1\columnwidth]{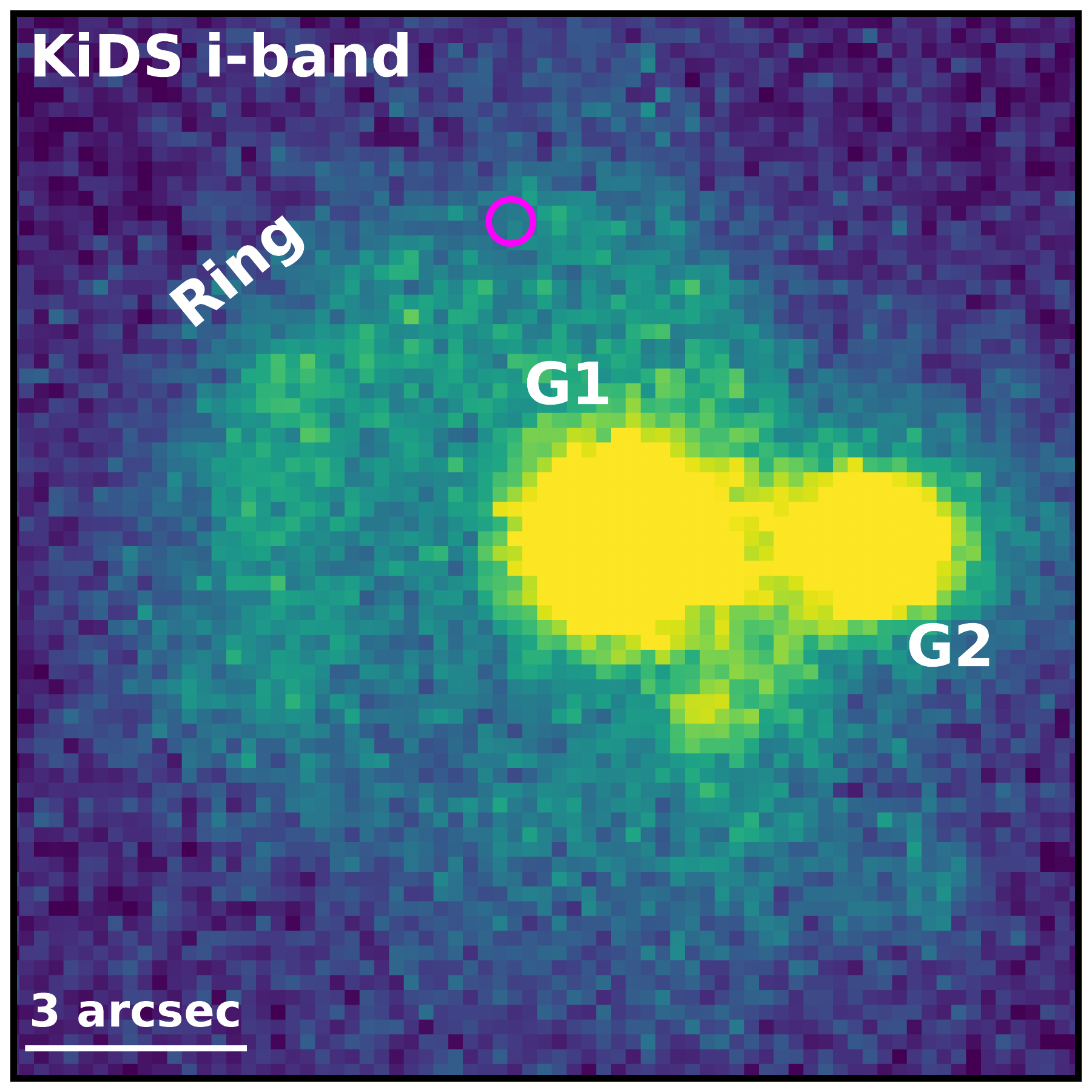}
    \hspace{-0.25cm}
    \includegraphics[width=1\columnwidth]{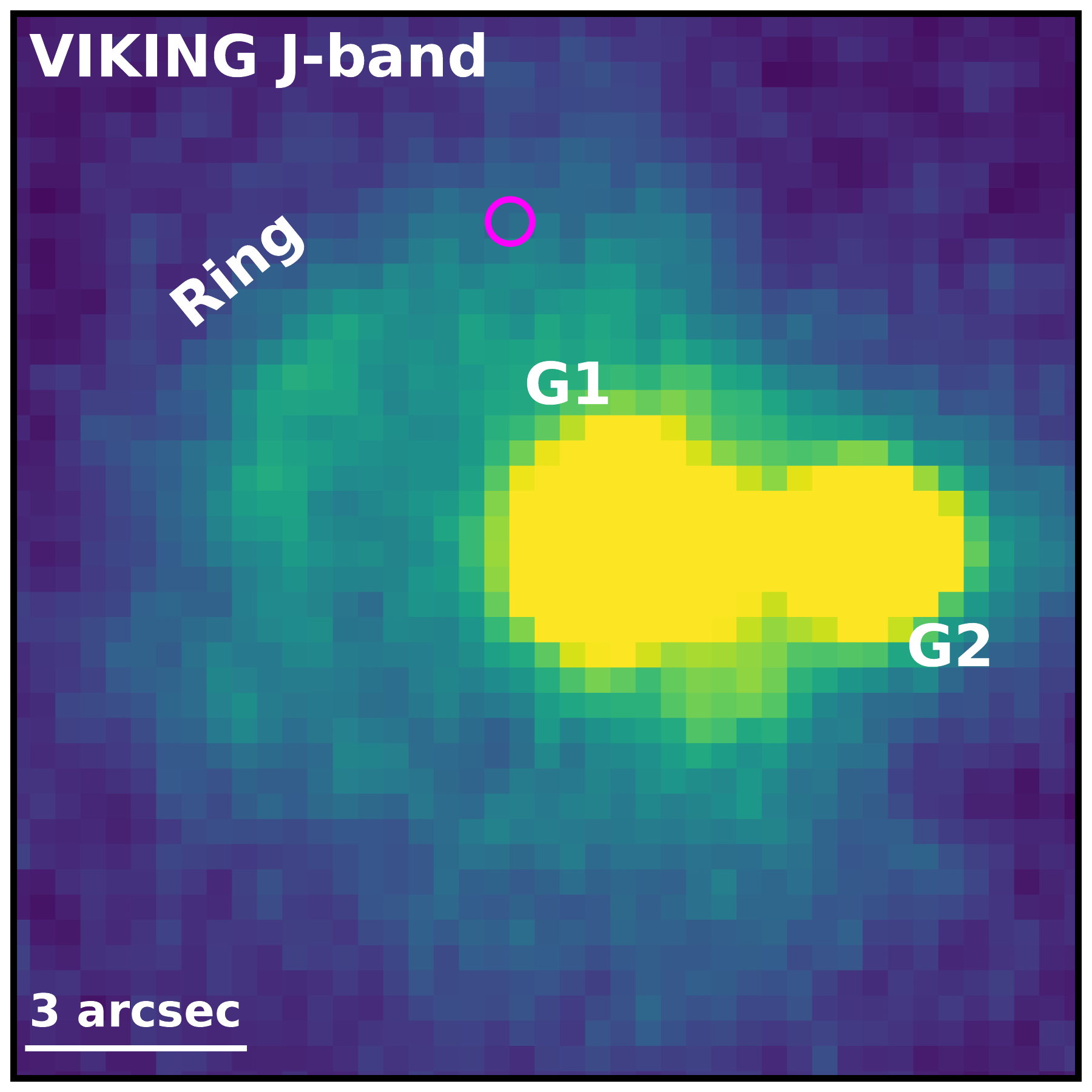}
    
        \caption{Individual band archival images from the KiDS and VIKING surveys of the collisional ring complex.}
    \label{fig:DeepImageAll1}
\end{figure*}

We present single band pre-explosion imaging of the field in Figure \ref{fig:DeepImageAll1}.

\section{Morphological Modeling of the Source Complex}

Due to the brightness of the galaxies G1 and G2 it is difficult to find a image scale that shows the full range of ring features without significant visual blending. Therefore, we modeled G1 and G2 to subtract them from the image so that the ring can be more easily resolved. This process also allows the possible tidal feature to the West of G2 to be better revealed. We focused on the KiDS $r$-band imaging as it has the highest quality seeing \citep{2013Msngr.154...44D}. We used \texttt{photutils} \citep{Bradley2024} to derive an empirical point spread function (PSF) from bright, isolated field stars and then applied this as a convolution kernel to a 2D Sérsic profile from \texttt{astropy} \citep{Astropy2018}. We performed a fit using non-linear least squares optimization from \texttt{scipy} \citep{scipy}. The sky level and background variations were estimated from a source free annulus around the ring, and pixel uncertainties were derived from the KiDS weight map. We found that both G1 and G2 required the combination of two Sérsic components (per galaxy) to model a small bulge and the larger halo. 

\begin{figure*}
    \centering
  \includegraphics[width=2\columnwidth]{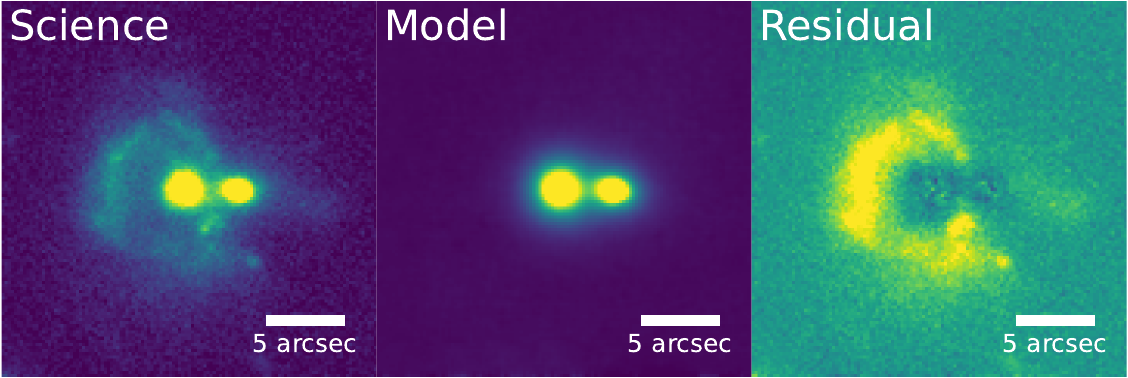}
    
        \caption{Morphological analysis of archival KiDS $r$-band imaging. North is up and East is to the left.}
    \label{fig:ResidImage}
\end{figure*}

\section{Additional Spectral Figures}

\begin{figure}
    \centering
    \includegraphics[width=1\columnwidth]{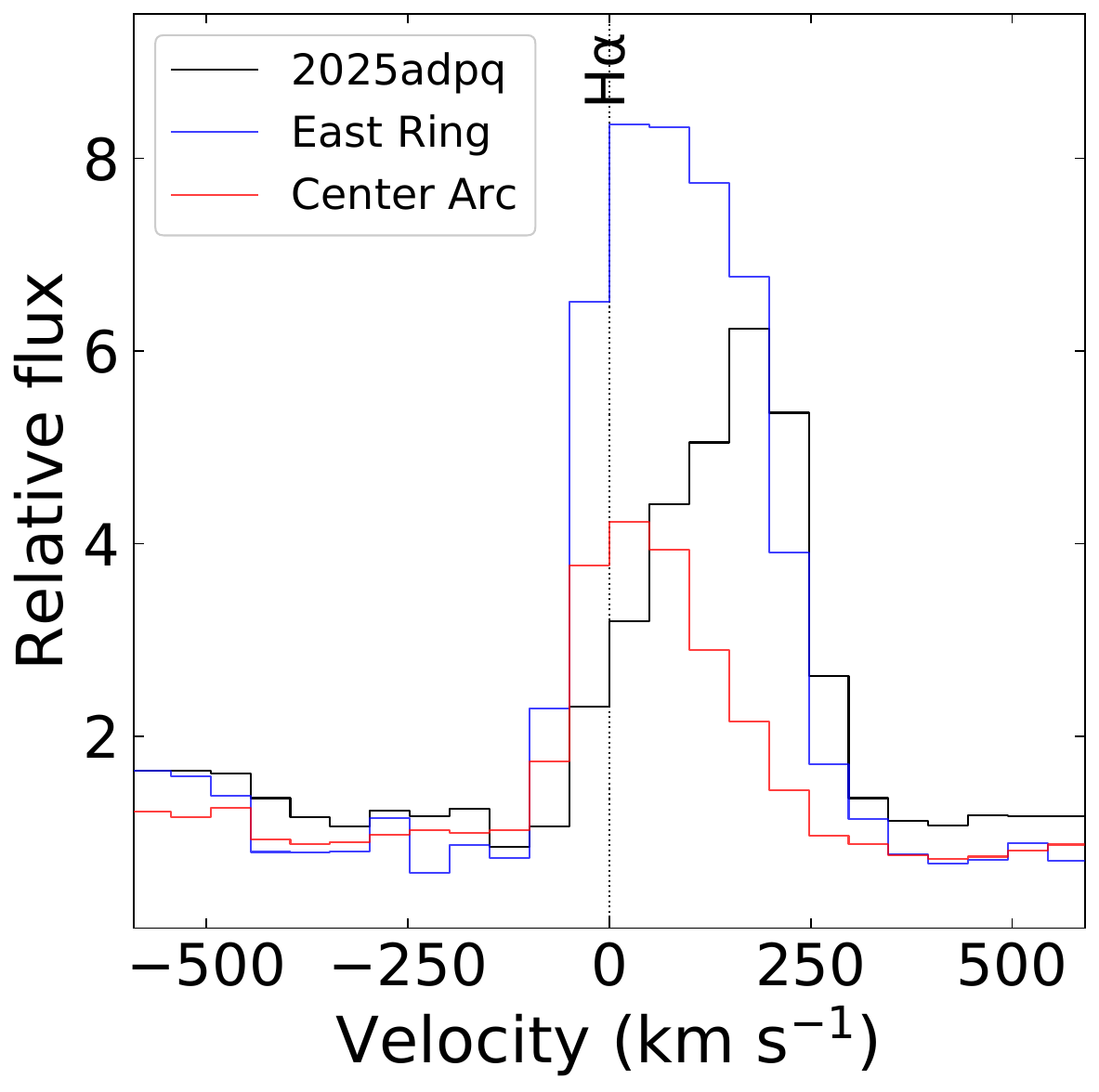}
    \caption{Zoom on the H$\alpha$ region showing velocity shifts with respect to rest, assuming the redshift of the host galaxy G1 at $z$\,$\approx$\,$0.1537$. There is a clear velocity shift along the ring. We note that the redshift of the ring at this location is $z$\,$\approx$\,$0.1540$, slightly redshifted with respect to G1.}
    \label{fig:GemHalphaZoom}
\end{figure}

\begin{figure*}
    \centering
    \includegraphics[width=1\columnwidth]{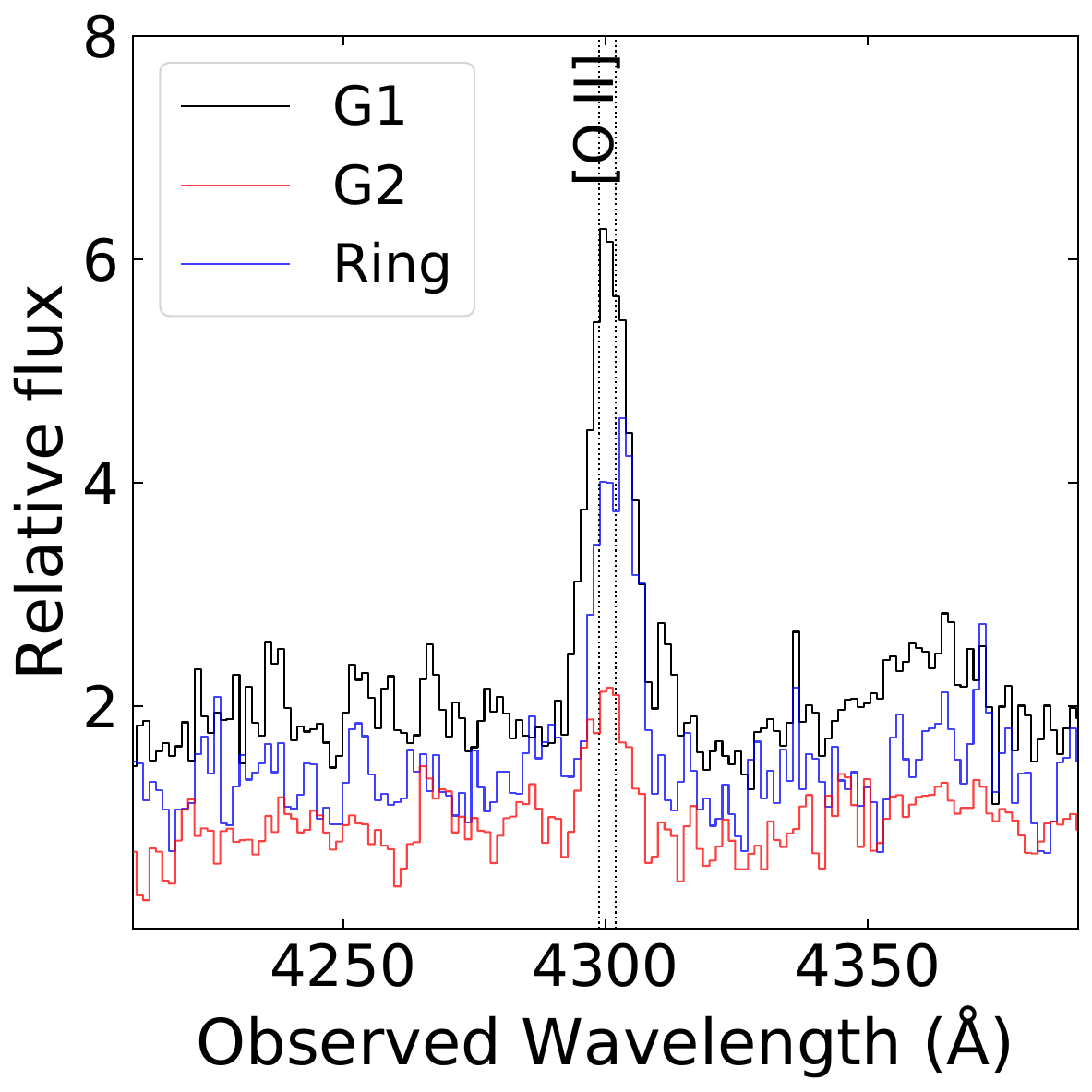}
    \includegraphics[width=1\columnwidth]{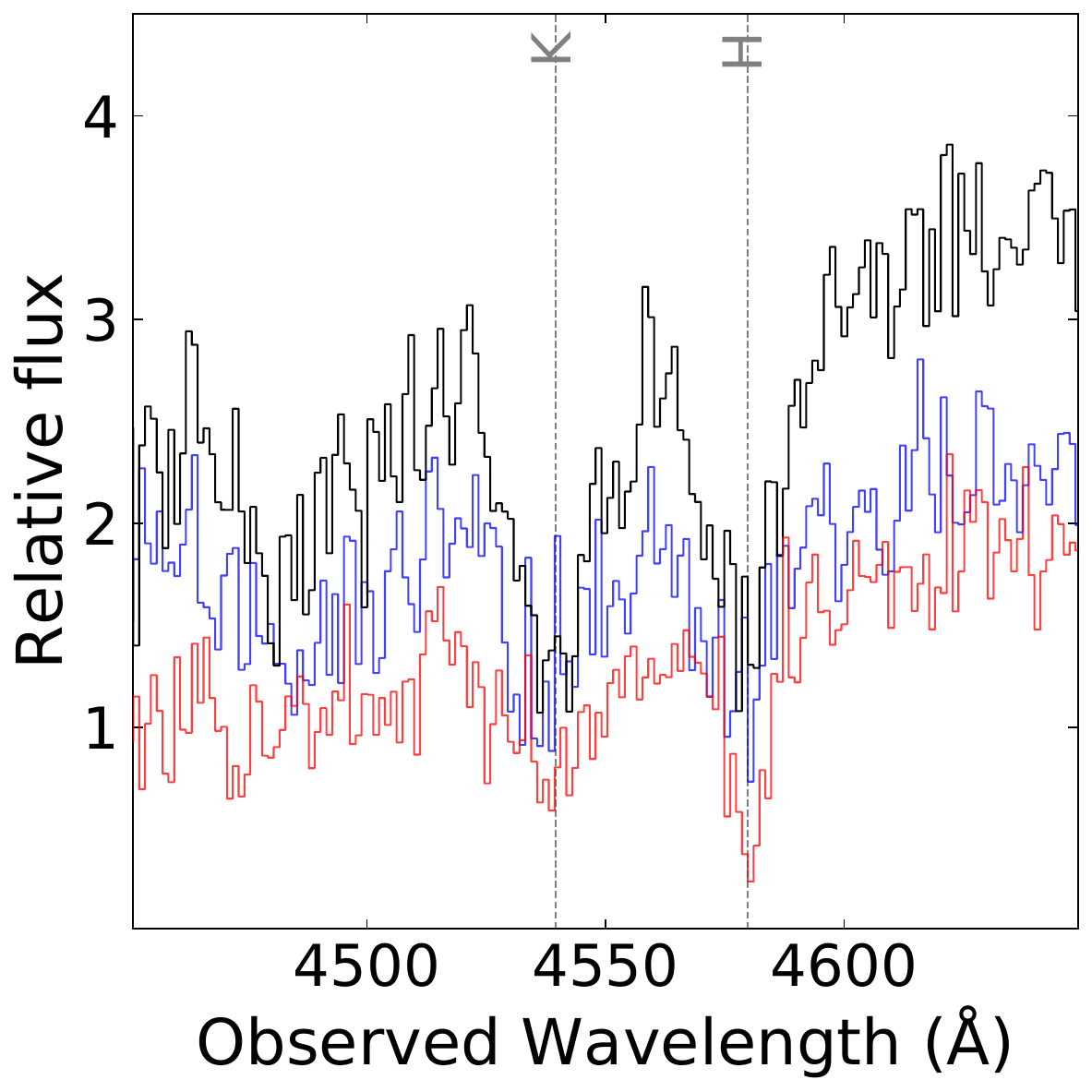}
    \caption{Zoom on the SALT/RSS spectra of G1, G2, and the ring showing the [O\,\textsc{ii}] doublet (left panel). and the Calcium H and K absorption lines (right panel). The sources all lie at a common redshift ($z$\,$=$\,$0.1537$) with small (e.g. hundred of km s$^{-1}$) differences in velocity.}
    \label{fig:SALTzooms}
\end{figure*}

We provide a zoom in on relevant spectral features identified by Gemini and SALT in Figures \ref{fig:GemHalphaZoom} and \ref{fig:SALTzooms}, respectively. 

\section{Additional Results of Galaxy SED Modeling}

\begin{figure*}
    \centering
    \includegraphics[width=2\columnwidth]{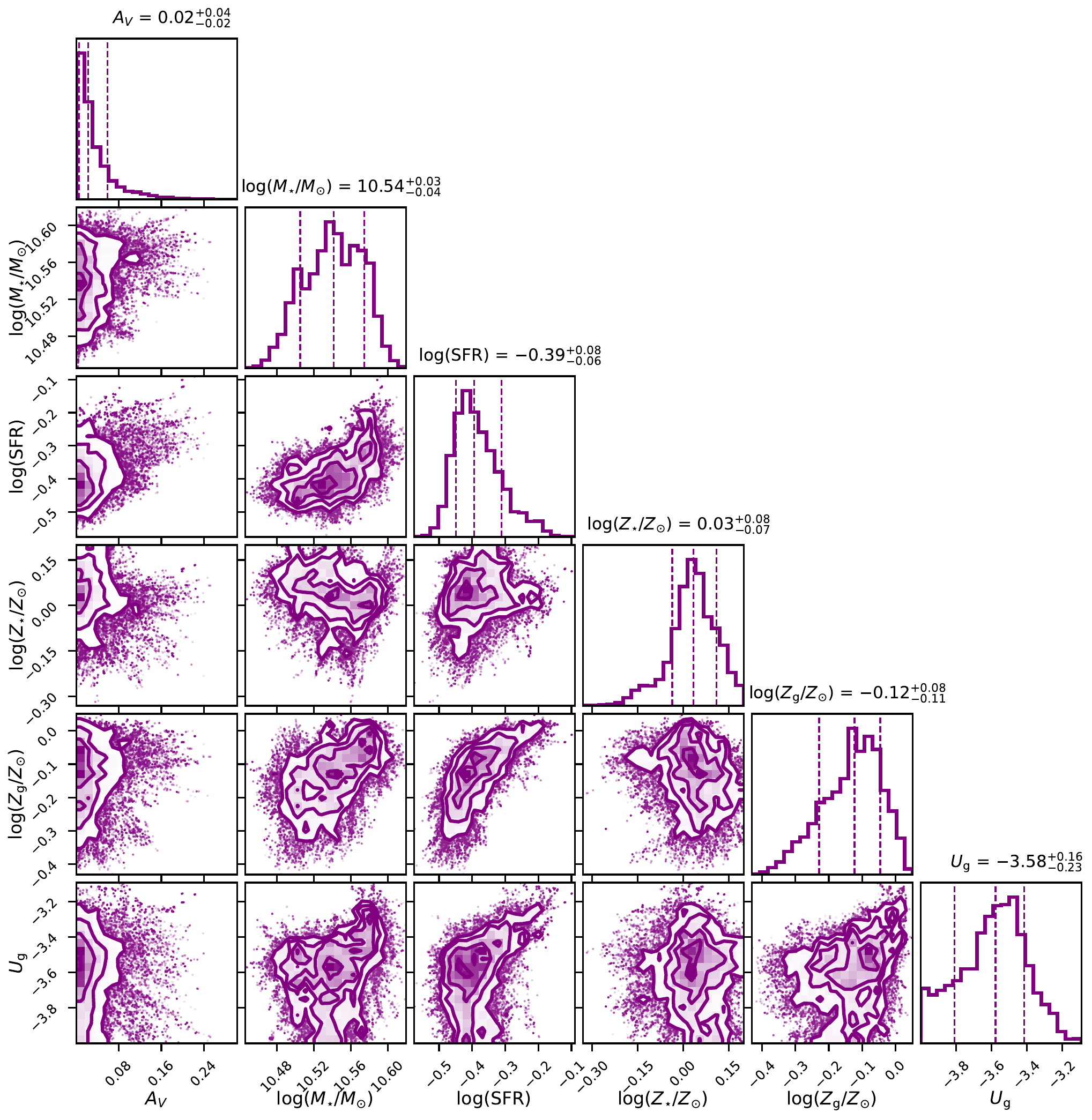}
    \caption{Corner plot showing the results of our \texttt{prospector} modeling for G1.}
    \label{fig:prospectorCorner1}
\end{figure*}

\begin{figure*}
    \centering
    \includegraphics[width=2\columnwidth]{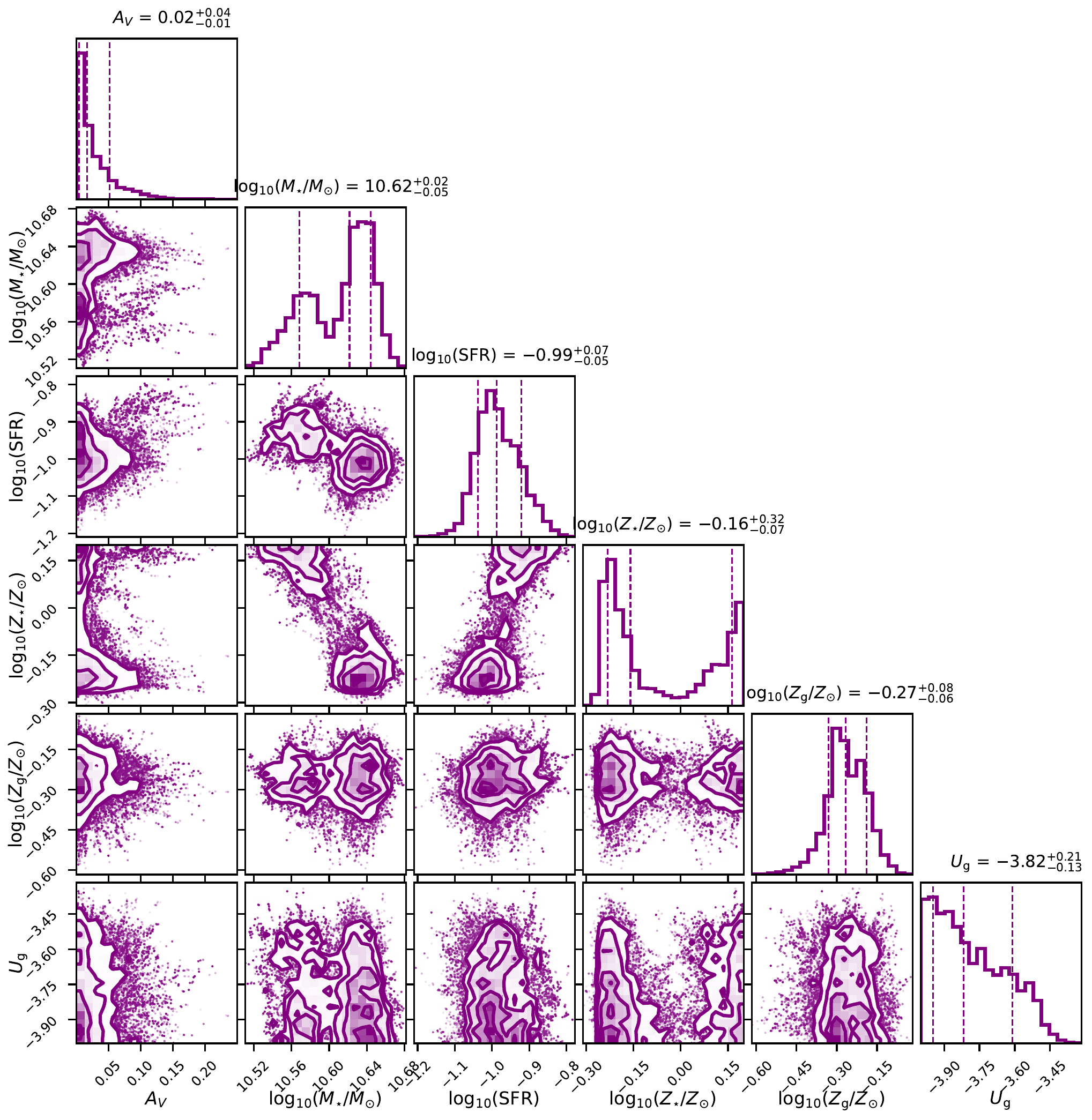}
    \caption{Corner plot showing the results of our \texttt{prospector} modeling for G2.}
    \label{fig:prospectorCorner2}
\end{figure*}

\begin{figure*}
    \centering
    \includegraphics[width=1\columnwidth]{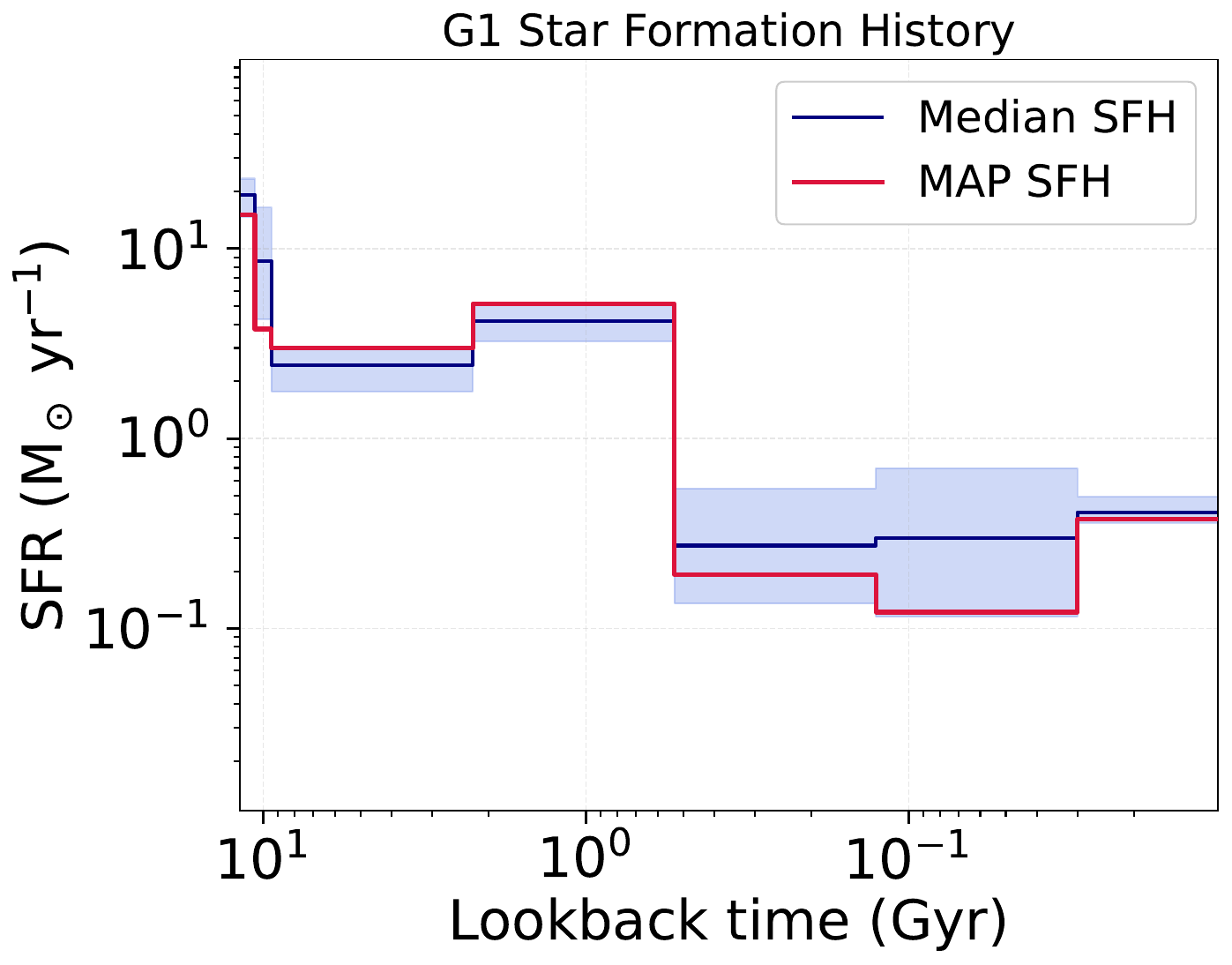}
    \includegraphics[width=1\columnwidth]{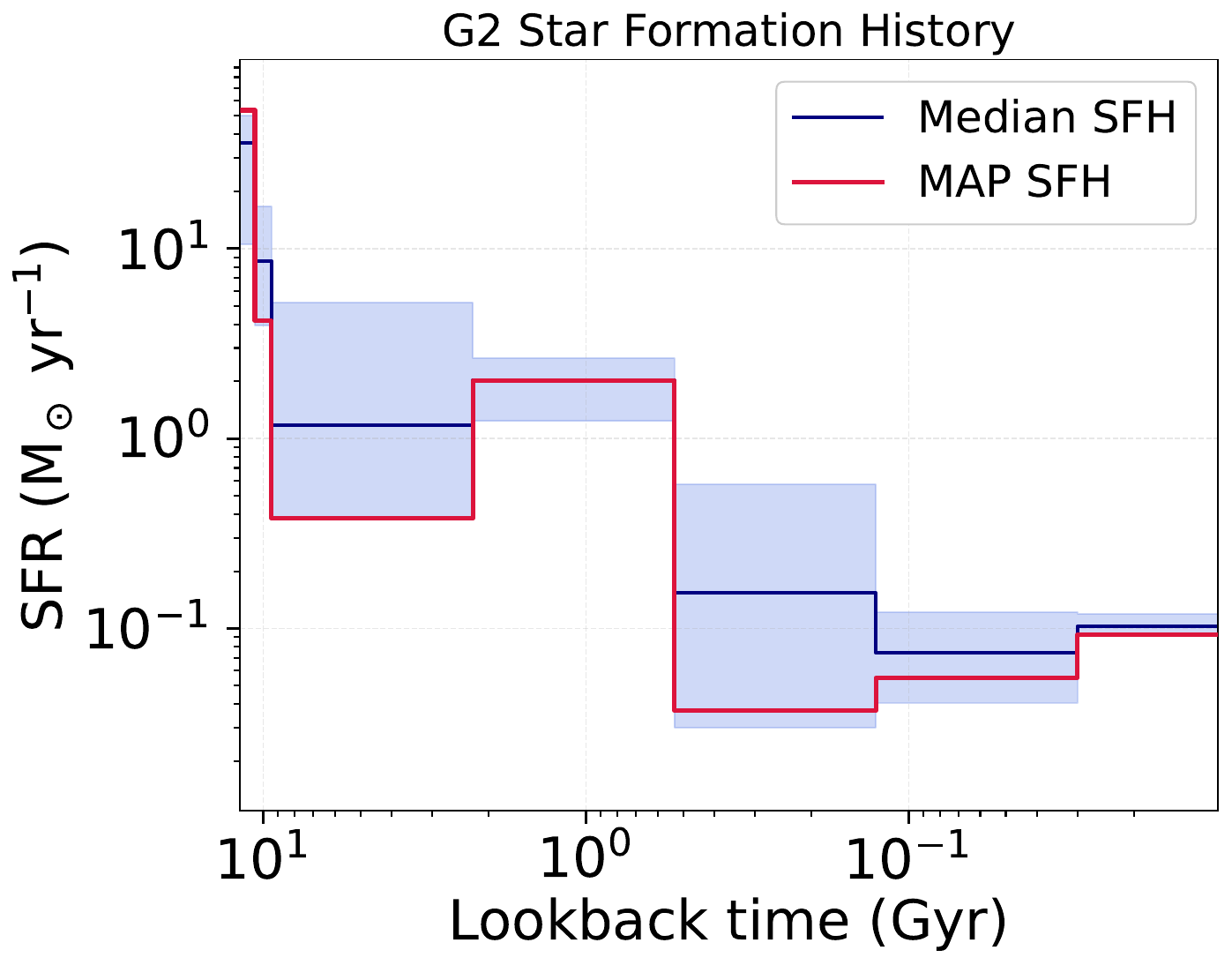}
    \caption{Star formation histories of G1 (left) and G2 (right) as derived with \texttt{prospector} \citep{Leja2019nonpar}. The nonparametetric SFH were derived in 7 age bins. The median SFH (blue) and $1\sigma$ confidence (light blue shaded) are shown. The best fit (MAP) SFH is shown in red. }
    \label{fig:SFH}
\end{figure*}

The results of our galaxy SED modeling (\S \ref{sec:prospector}) as shown in Figure \ref{fig:prospectorCorner1} for G1 and Figure \ref{fig:prospectorCorner2} for G2. Figure \ref{fig:SFH} shows the derived star formation history of both galaxies.

\section{Results of Supernova Lightcurve Modeling}
\label{sec:lcfitcorner}

\begin{figure*}
    \centering
    \includegraphics[width=2\columnwidth]{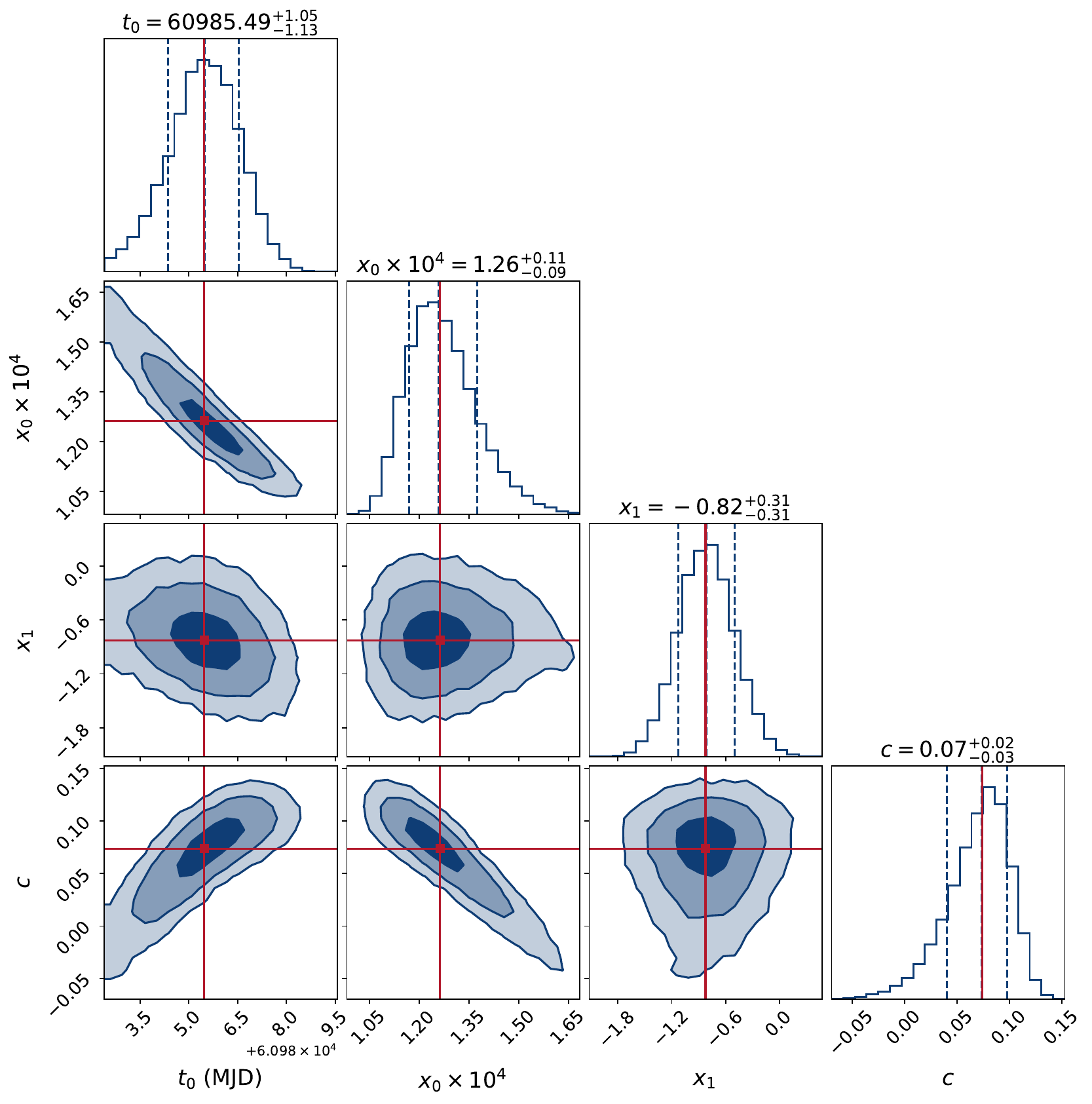}
    \caption{Corner plot showing the results of our SALT2 modeling of the supernova lightcurve.}
    \label{fig:lcfitcorner}
\end{figure*}

In Figure \ref{fig:lcfitcorner}, we display the corner plot showing the posterior distributions for each parameter based on our SALT2 modeling of the supernova lightcurve.

\bibliography{bib}{}
\bibliographystyle{aasjournal}

\end{document}